\documentclass[twocolumn,prx,hyperref,footinbib,longbibliography]{revtex4-1}
\usepackage{graphicx}
\usepackage{dcolumn}
\usepackage{bm}
\usepackage{physics}
\usepackage{slashed}
\usepackage{amsmath,amssymb}
\usepackage{array} 
\usepackage{float}
\usepackage{multirow}
\usepackage{makecell}

\usepackage{diagbox, eqparbox, hhline}
\newcolumntype{P}[1]{>{\centering\arraybackslash}p{#1}}

\usepackage{hyperref}
\usepackage{mathtools}

\usepackage[normalem]{ulem}  
\usepackage{natbib}
\usepackage{threeparttable}
\usepackage{tabularx}
\usepackage{xcolor}
\usepackage{adjustbox} 
\usepackage{flafter} 
\usepackage{booktabs}

\begin{document}

\title{Precise high-fidelity electron-nuclear spin entangling gates in NV centers via hybrid dynamical decoupling sequences}

\author{Wenzheng Dong}
\email{dongwz@vt.edu}
\author{F. A. Calderon-Vargas}
\email{f.calderon@vt.edu}
\author{Sophia  E.  Economou}
\email{economou@vt.edu}
 \affiliation{ Department of Physics, Virginia Tech, Blacksburg, Virginia 24061, USA}

\begin{abstract}
Color centers in solids, such as the nitrogen-vacancy center in diamond, offer well-protected and well-controlled localized electron spins that can be employed in various quantum technologies. Moreover, the long coherence time of the surrounding spinful nuclei can enable a robust quantum register controlled through the color center. We design pulse sequence protocols that drive the electron spin to generate robust entangling gates with these nuclear memory qubits. We find that compared to using Carr-Purcell-Meiboom-Gill (CPMG) alone, Uhrig decoupling sequence and hybrid protocols composed of CPMG and Uhrig sequences improve these entangling gates in terms of fidelity, spin control range, and spin selectivity. We provide analytical expressions for the sequence protocols and also show numerically the efficacy of our method on nitrogen-vacancy centers in diamond. Our results are broadly applicable to color centers weakly coupled to a small number of nuclear spin qubits.

\end{abstract}

\maketitle

\section{Introduction}
Color centers in solids provide well-isolated local electron spins with long coherence times \cite{BalasubramanianNatMat2009,KennedyAPL2003,WidmannNatureMat2014,RadulaskiNanoLetters2017,Bar-GillNatCom2013}, high fidelity manipulation \cite{NagyNatComm2019,KoehlNature2011,RobledNature2011,Dolde2014}, and typically a small set of surrounding nuclear spins which can act as a quantum register. As a result, color centers are promising platforms to realize quantum technologies \cite{AwschalomNature2018,Atature2018,SiminPRX2016,HumphreysNature2018,SukachevPRL2017,RendlerNatComm2017,SchlipfeScience2017,LaydenPRL2020}, including quantum sensing \cite{AslamScience2017,AnisimovSciRep2016,ShiScience2015,LeePRB2015,AbobeihNature2019,Degen.Cappellaro.RMP2017}, quantum communication \cite{HensenNature2015,BernienNature2013,NguyenPRL2019}, and quantum computing \cite{CramerNatCom2016,WuNPJ2019,BradleyPRX2019}. Among the most actively studied color center platforms  are nitrogen-vacancy (NV) centers  \cite{DohertyNJP2011,vanDamPRL2019,Taminiau:2014aa,ChoiPRL2017} and  silicon-vacancy (SiV)  \cite{MeesalaPRB2018,BeckerPRL2018,PingaultPRL2014,SunShuoPRL2018} in diamond, and divacancy \cite{SeoNatComm2016,AbramPRL2015,KoehlNature2011,ChristleNatMat2014,FalkNatCom2013,alex2020entanglement} and monovacancy centers in silicon carbide (SiC) \cite{NiethammerNatComm2019,FuchsNatCom2015,DaniilLukinNatPhoto2019,NagyPRAP2018,SoykalPRB2016,DongPRB2019}.

Even though the long coherence time of the nuclear spins makes them promising candidates for quantum memory \cite{AbobeihNature2019,MoChenNJP2018}, entanglement purification \cite{DurPRL2003,FujiiPRA2009}, and quantum nodes \cite{Nickerson2013} in quantum computing and communication, one of the main challenges is that the interactions between the electronic spin and the nuclear spin memory qubits are always on. As a result, controlling the system is not straightforward and the electron  spin coherence is hampered by the spinful isotopic nuclei in the host crystal.
In this regard, the Delft group has introduced and successfully demonstrated a clever way to address both of these issues: they have shown that appropriately chosen Carr-Purcell-Meiboom-Gill-like (CPMG-like)  dynamical decoupling sequences \cite{deLange60,GULLION1990479} applied on the electron spin not only protect it from the nuclear spin bath but also allow it to selectively control target nuclear spins via the hyperfine interaction \cite{TaminiauPRL2012,Taminiau:2014aa,CramerNatCom2016}. This novel use of CPMG-like sequences to implement two-qubit gates along with the vast repertoire of alternative dynamical decoupling sequences~\cite{Viola1999,Khodjasteh2005,UhrigPRL2007,UhrigPRL2009}, opens the question of whether there are even better nuclear spin control protocols \textcolor{black}{that can be achieved through the combination of always-on interactions and drive of the electron spin.}

In this paper, we introduce new, advantageous ways of selective, fast, and high-fidelity electron-nuclear spin entangling gates through pulse sequences acting on the electronic spin. We specifically focus on Uhrig dynamical decoupling (UDD) sequences \cite{UhrigPRL2007} and on \emph{hybrid} protocols based on a combination of CPMG-like \cite{deLange60,GULLION1990479} and UDD sequences. Our approach yields precise nuclear spin manipulation and good electron spin coherence protection. We find that, for a wide range of magnetic fields, the hybrid sequences provide fast electron-nuclear two-qubit gates with higher fidelity than what would be obtained by only using CPMG or UDD alone. Moreover, in contrast to other sequences, UDD provides high spin selectivity without significantly increasing the overall gate time. This facilitates the precise control of nuclear spins with similar hyperfine interaction strengths. Interestingly, and contrary to what one may conclude based on prior literature, we find that UDD provides better electron spin coherence protection compared to CPMG in the parameter regime that accomplishes high spin selectivity. 
We test our sequences numerically on an NV center in diamond, using system parameters from experiment \cite{TaminiauPRL2012}. 
Our protocol is general, and can thus be applied to similar platforms--e.g., in divacancy centers in SiC, where the CPMG protocol was recently used to control nuclear spins \cite{alex2020entanglement}--after straightforward modifications. 

\begin{figure} [h]
\includegraphics[width=\linewidth]{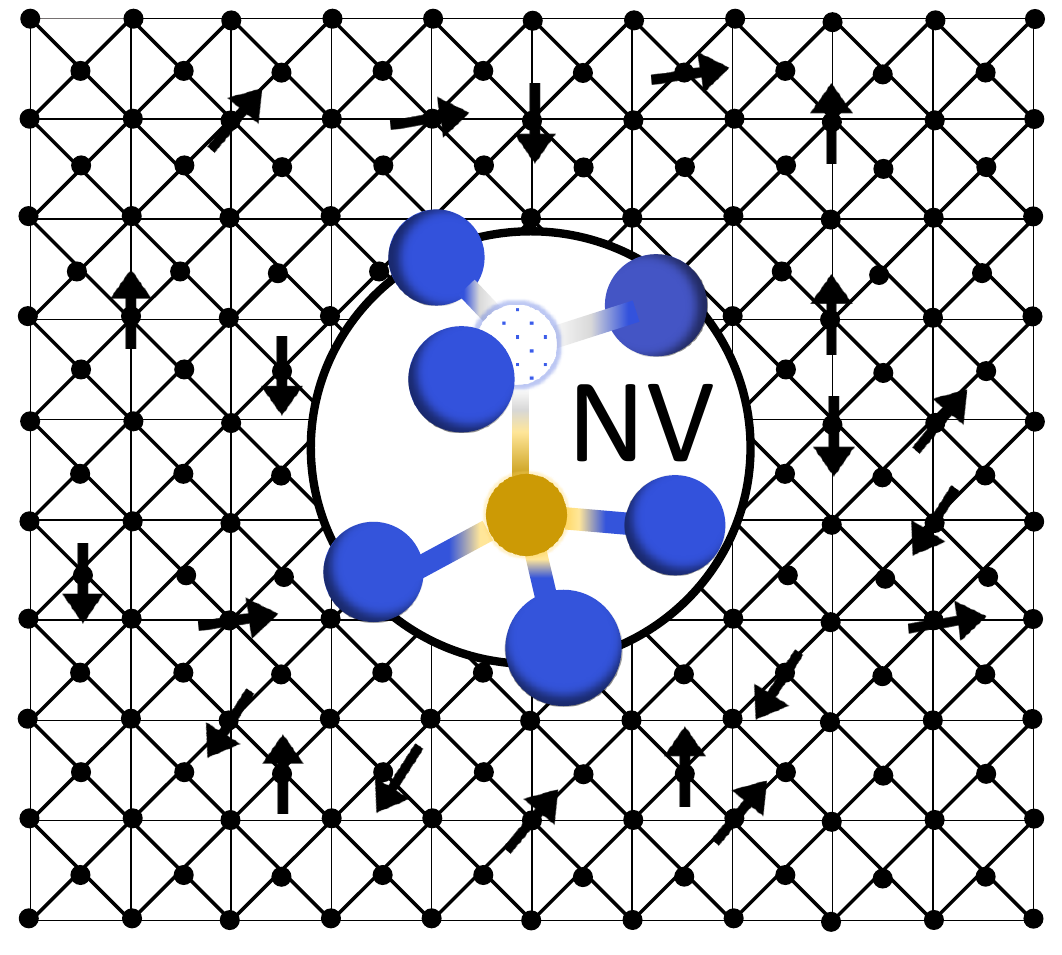}
\caption{Sketch of an NV center in diamond (top view of an extended (100) plane). The  defect consists of a nitrogen substitute (yellow) and a vacancy (dotted white) at the neighboring site, and features a localized $S=1$ electronic spin. Individual nuclear spins (black arrows) are coupled via dipolar hyperfine interaction to the NV electronic spin.}
\label{fig:control_nspins_illustration}
\end{figure} 

The remainder of this paper is organized as follows. In Sec.~\ref{sec:II Modeling the system Hamiltonian}, we present the system's Hamiltonian. In Sec.~\ref{sec:III CPMG-like sequences inducing quantum gates on nuclear spins}, we review the use of CPMG-like sequences to control weakly coupled nuclear spins in NV centers. In Sec.~\ref{sec:IV Controlling nuclear spins with UDD sequences}, we introduce Uhrig dynamical decoupling sequence and investigate its performance in terms of fidelity, selectivity, and electron spin coherence protection. In Sec.~\ref{sec:V Controlling nuclear spins with hybrid sequences}, we present our new hybrid sequences for electron-nuclear spin entangling gates in NV centers, showing their versatility and overall performance. We conclude in Sec.~\ref{sec:Conclusions}.

\section{Modeling the system Hamiltonian}\label{sec:II Modeling the system Hamiltonian}
The geometric structure of the NV center, a nitrogen substitute and a neighboring carbon vacancy, is illustrated in Fig. \ref{fig:control_nspins_illustration}. 
The system under consideration is formed by the central electron spin and several weakly coupled nuclear spins (carbon isotope ${}^{13}_{}\text{C}^{}_{}$, with natural abundance of $1.1\%$) that we aim to control.
In the presence of an external magnetic field ($B$ field) applied along the $z$-axis, the total  Hamiltonian of the  system  is: 
\begin{equation}
\label {eq:total_Hamiltonian}
\begin{aligned}
H_{\text{total}}=H_E+H_{\text{bath}}+H_{\text{int}}, \\
\end{aligned}
\end{equation}
where $H_{E}$ is the Hamiltonian for the $S=1$ electron spin, $H_{\text{int}}$ is the hyperfine interaction between the central electron spin and all the nuclear spins in the system, each with spin $I=1/2$, and $H_{\text{bath}}$ is the nuclear spin bath Hamiltonian. Note that we  assume that the nuclear spins do not mutually interact, and thus $H_{\text{bath}}$ is reduced to the sum of  \textcolor{black}{Zeeman terms of the nuclear spins} (see Appendix~\ref{app: B - System Hamiltonian} for more details on the total Hamiltonian).
\textcolor{black}{Since the NV center zero-field splitting ($\sim 2.87$ GHz) is much larger than the typical hyperfine interaction strength ($<$MHz), it is highly unlikely that an electron spin-flip process occurs. Therefore, the transverse components of the electron spin can be safely neglected and only the $S_z$ term is kept}~\cite{ZhaoNanPRB2008,TaminiauPRL2012}.

It is convenient to start the analysis with the simplest case where the electron spin is only interacting with a single nuclear spin.  Accordingly, their Hamiltonian in the interaction picture given by $H_{E}$ is simply:
\begin{equation}
\label {eq:single_Hamiltonian}
\begin{aligned}
\tilde{H}=&\omega_{L}  {I_z} +\sum_{m_s}\sum_{i\in\{x,y,z\}} m_s\ket{m_s}\bra{m_s} \otimes \mathbb{A}_{z,i} \cdot{I}_{i}  \\ 
=& \omega_L I_z + S_z \otimes (A_{\perp} I_x+ A_{\parallel} I_z), \\
\end{aligned}
\end{equation}
where $\omega_{L}$  is the nuclear spin Larmor frequency,  $I_i$ is the cartesian component ($i=x,y,z$) of the nuclear spin operator, and $m_s$ is the magnetic spin quantum number of the electron, which can be equal to -1, 0, or 1. \textcolor{black}{For a sufficiently strong magnetic field, the $m_s=\pm 1$ states are well separated in frequency, allowing us to treat the spin system as an effective two-level system with the $m_s=0$ state and one of the $m_s=\pm 1$ states forming the two qubit levels.} 
We follow the notation of Ref. \cite{TaminiauPRL2012} and define $\ket{m_s=-1}=\ket{1}$ and $\ket{m_s=0}=\ket{0}$ to encode the qubit and introduce the $z$-component of the pseudo-spin operator $S_z=0\ket{0}\bra{0}-\ket{1}\bra{1}$, where, for the sake of simplicity, we set $\hbar=1$. 
Note that due to the diagonal form of the electron spin operator, the elements of the hyperfine interaction tensor, $\mathbb{A}_{j,i}$, are nonzero only for $j=z$ and $i=x,y,z$ (see Appendix~\ref{app: B - System Hamiltonian}).
Moreover, the hyperfine interaction elements can be reduced to parallel and perpendicular components with respect to the $z$-axis, i.e. $A_{\parallel}$ and $A_{\perp}$, respectively, by rotating the $x-y$ plane. Therefore, Eq.~\eqref{eq:single_Hamiltonian} can be expressed as:
\begin{equation}\label{eq:00h_0+11h_1 Hamiltonian}
\begin{aligned}
\tilde{H}=&\ket{0}\bra{0} \omega_L I_z +\ket{1}\bra{1} [(\omega_L-A_{\parallel})I_z- A_{\perp} I_x  ] \\
=& \ket{0}\bra{0} h_0+\ket{1}\bra{1}h_1, 
\end{aligned}
\end{equation}
where $h_0=\omega_L I_z $ and $h_1=(\omega_L-A_{\parallel})I_z- A_{\perp} I_x$. 
Similarly, for  multiple  nuclear spins the total Hamiltonian can be written as:
\begin{equation}
\label{eq:general_H_E-n}
\tilde{H}_{\text{total}}=\ket{0}\bra{0}\otimes \sum^n_{i=1}  h^{(i)}_0 +\ket{1}\bra{1} \otimes \sum^n_{i=1}  h^{(i)}_1
\end{equation}
where $n$ is the number of nuclear spins and $h^{(i)}_0(h^{(i)}_1)$ is a multi-nuclear operator consisting of the tensor product of $h_0(h_1)$, which acts on the $i$-th nuclear spin, and the identity operator acting on the remaining nuclear spins.

\section{Review of CPMG-like quantum gates }\label{sec:III CPMG-like sequences inducing quantum gates on nuclear spins}

To manipulate nuclear spins surrounding the central electron spin, Taminiau \textit{et al.} used CPMG-like dynamical decoupling sequences, namely the XY8 sequence~\cite{TaminiauPRL2012,deLange60,GULLION1990479} to manipulate several nuclear spins weakly coupled to the central electron spin of a NV center in diamond. This sequence is applied on the central electron spin to decouple it from the surrounding nuclear/electron spin bath, thus extending its coherence time. At the same time, Ref. \cite{TaminiauPRL2012} has shown that it is possible to induce conditional rotations on a target nuclear spin by tuning the sequence's inter-pulse delay time to satisfy a resonance condition determined by the hyperfine interaction between the electron spin and the target nuclear spin. \textcolor{black}{The use of periodic pulses in other systems involving coupled electron and nuclear spins, such as quantum dots \cite{beugeling2017influence,frohling2019fourth,kleinjohann2018magnetic,frohling2018nuclear,beugeling2017influence,kleinjohann2018magnetic,jaschke2017nonequilibrium}, produces similar resonance conditions.}

The CPMG-like dynamical decoupling sequence used in Ref.~\onlinecite{TaminiauPRL2012}, hereinafter referred to as simply CPMG sequence, consists of a train of pulses with the basic decoupling unit being $(\tau-\pi-2\tau-\pi-\tau)^N$, where $\pi$-pulses ($\pi$ rotations about the $x$-axis and $y$-axis in an alternating fashion) are applied to the electron spin, separated by a $2\tau$ delay time, and $N$ is the total number of basic decoupling units in the sequence. 
In the following analysis we use the total time $t\equiv4\tau$ of the basic decoupling unit instead of the inter-pulse distance $2\tau$ to derive the resonance condition, as it is more convenient for comparing it to other types of dynamical decoupling sequences.

Now, if we apply a single basic CPMG unit ($N=1$) to the system formed by the central electron spin and $n$ nuclear spins, Eq.~\eqref{eq:general_H_E-n}, the evolution operator would be:
\begin{equation}
\label{eq:total_Evol_under_cpmg}
U_{ }=\ket{0}\bra{0} \otimes \prod^n_k V^{(k)}_0 +\ket{1}\bra{1} \otimes \prod^n_k V^{(k)}_1, 
\end{equation}
where the electron-spin-state-dependent evolution operators acting on the $k-$th nuclear spin are $V^{(k)}_0= I^{(1)} \otimes ... \otimes e^{-i h^{(k)}_0 \tau}e^{-i h^{(k)}_1 2\tau}e^{-i h^{(k)}_0 \tau}\otimes ... \otimes I^{(n)}$  and $V^{(k)}_1= I^{(1)} \otimes ...\otimes e^{-i h^{(k)}_1 \tau}e^{-i h^{(k)}_0 2\tau}e^{-i h^{(k)}_1 \tau} \otimes... \otimes I^{(n)}$. The full CPMG sequence contains $N$ copies of the basic unit, and thus the total evolution operator is $U^N$. 
Taminiau et al. \cite{TaminiauPRL2012} demonstrated that in a strong magnetic field and for a weakly coupled nuclear spin ($\omega_L\gg A_{\perp},A_{\parallel}$), whenever the total basic unit time $t$ (or similarly the inter-pulse distance $2\tau$) satisfies a resonance condition the electron and target nuclear spins become coupled and the latter \textcolor{black}{undergoes a rotation that is conditional on the electron spin state}. 
Alternatively, when the basic unit time $t$ is not on resonance with any target nuclear spin, the nuclear spins are decoupled from the electron spin and they just  unconditionally rotate about an axis and rotation angle determined by how far from resonance $t$ is for each nuclear spin. 

\begin{figure} [h]
\includegraphics[width=\linewidth]{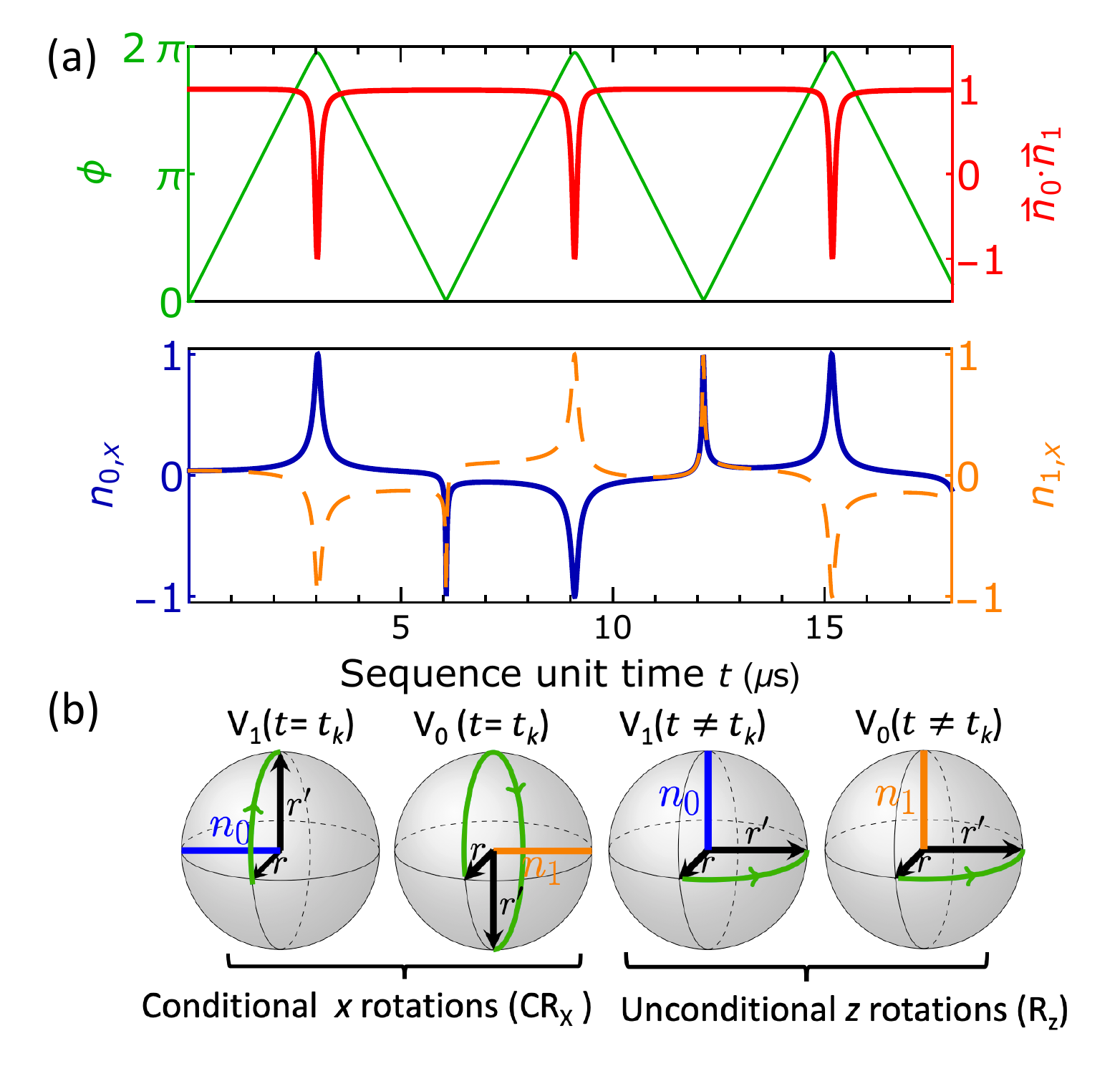} 
\caption{CPMG control of a single nuclear spin. (a) The axes of nuclear spin rotation conditional on the electronic spin, their dot product, and the angle of rotation as functions of the sequence time $t$. The red curve represents the dot product of the rotation axes, $\vec{n}_0 \cdot \vec{n}_1$, of  the electron spin state-dependent evolution operators $V_0$   and $V_1$,  where the periodic dips ($\vec{n}_0 \cdot \vec{n}_1=-1$) indicate the conditional rotations with opposite rotational axes and the flat portions of the curve ($\vec{n}_0 \cdot \vec{n}_1=1$) indicate unconditional rotations. The green curve shows the rotation angle. At the resonant points the rotation angles plotted here are close to $2\pi$, which indicates small effective rotation angles. The blue (orange dashed) curve represents the $x$-direction projection of the rotation axis $\vec{n}_0$ ($\vec{n}_1$). The peaks/dips in the blue/orange curve indicate $x$ rotations, of which the ones that are synchronous with the red curve dips correspond to conditional $x$ rotations ($CR_X$) and the rest denote unconditional $x$ rotations ($R_X$). At any other values of $t$, the nuclear spin rotates along the $z$-axis unconditionally. In these simulations we used ${A_{\parallel}}/{2\pi}=30.6$ kHz, ${A_{\perp}}/{2\pi}=25.7$ kHz and $\omega_L/2\pi=314$ kHz.  
(b) Qualitative illustration of conditional $x$ and unconditional $z$ rotations. }
\label{fig:geometric_representation}
\end{figure} 

To characterize the two-qubit gates emerging from the CPMG sequences acting on the electron spin, let's consider the simple case of a single nuclear spin interacting with the central electron spin. Following Ref. \cite{TaminiauPRL2012}, we can express the conditional evolution operators $V_0$ and $V_1$, Eq.~\eqref{eq:total_Evol_under_cpmg}, as $V_0=\exp [-i\phi (\vec{I}\cdot\vec{n}_0)]$ and $V_1=\exp [-i\phi (\vec{I}\cdot\vec{n}_1)]$, respectively. Here $\phi$ is the rotation angle, $\vec{n}_{|m_s|}$ is the rotation axis that depends on the electron's initial state $m_s=0$ or $m_s=-1$ , and $\vec{I}$ is the nuclear spin operator. As shown in Fig.~\ref{fig:geometric_representation}, the inner product of the rotation axes $\vec{n}_0\cdot\vec{n}_1$ indicates whether the nuclear spin rotation induced by the CPMG sequence is conditional ($\vec{n}_0\cdot\vec{n}_1=-1$) or unconditional ($\vec{n}_0\cdot\vec{n}_1=1$). The conditional rotations are controlled-$R_{\pm X}(\phi)$ ($CR_X(\phi)$), i.e. $x$-rotations by an angle $\phi$ with a direction that depends on the electron spin state, and the unconditional ones are simply nuclear spin rotations about the $x$-axis, $z$-axis, or an axis in between the previous two, that does not depend on the electron spin state.

In order to generate a $CR_X(\phi)$ gate, the CPMG unit time $t$ must satisfy a resonance condition determined by the Larmor frequency of the nuclear spin and by the hyperfine interaction between the target nuclear spin and the electron spin. \textcolor{black}{For a sufficiently strong magnetic field ($\omega_L \gg A_{\parallel},A_{\perp}$), the resonance condition can be found analytically~\cite{TaminiauPRL2012,Taminiau:2014aa,CramerNatCom2016}.} Accordingly, at resonance, the CPMG unit time $t^{\text{CPMG}}$ and the nuclear spin rotation angle $\phi^{\text{CPMG}}$ are~\cite{TaminiauPRL2012}:
\begin{equation}
\label{eq:conditional_rotation}
	t_k^{\text{CPMG}} \approx\frac{4(2k-1)\pi}{2\omega_L-A_{\parallel}},  \quad \quad   \quad     
	\phi^{\text{CPMG}} \approx 2\pi- \frac{2A_{\perp}}{\omega_L-A_{\parallel}}, \\
\end{equation}
where  $k$ is a positive integer. On the other hand, when $t$ is in the middle of two neighboring resonance values, $\frac{1}{2}(t_k^{\text{CPMG}}+t_{k+1}^{\text{CPMG}})$, the uncoupled nuclear spin rotates unconditionally about the $x$-axis through an angle $\tilde{\phi}^{\text{CPMG}}$, $R_X(\tilde{\phi})$. 
The CPMG unit time and rotation angle for the unconditional $R_X(\tilde{\phi})$ gate are:
\begin{equation}
\label{eq:unconditional_rotation}
	\tilde{t}_k^{\text{CPMG}}  \approx\frac{8 k\pi}{2\omega_L-A_{\parallel}},     \quad  \quad 
	 \tilde{\phi}_k^{\text{CPMG}}  \approx \frac{k\pi A_{\perp}A_{\parallel}}{(\omega_L-A_{\parallel})^2}. 
\end{equation}
Note that the rotation angle $\tilde{\phi}^{\text{CPMG}}$ does depend on $k$, in contrast to the resonance rotation angle $\phi^{\text{CPMG}}$ in Eq.~\eqref{eq:conditional_rotation}. These analytical expressions for the rotation angles as well as the analytical expressions for both $t_k^{\text{CPMG}}$ and $\tilde{t}_k^{\text{CPMG}}$ are approximations \cite{TaminiauPRL2012} and their accuracy is inversely proportional to $k$. Moreover, the integer term $k$ in Eqs.~(\ref{eq:conditional_rotation},~\ref{eq:unconditional_rotation}) is chosen to be as small as possible to avoid unnecessarily long sequences that may negatively affect the coherence protection of the electron spin. Note also that the relatively sharper peaks and dips corresponding to the $x$  projection of the unconditional rotation axis plot in Fig.~\ref{fig:geometric_representation} imply that the experimental timing precision required to implement an unconditional rotation is higher than the one required to implement a conditional rotation. That also suggests that the analytical approximation for $\tilde{t_k}^{\text{CPMG}}$ must be numerically optimized to increase its accuracy, and thus improve the resulting single-qubit gate fidelity \textcolor{black}{(the numerical optimization is implemented through minimization of $\vec{n}_0 \cdot \vec{n}_1$ by varying the time $t$)}. \textcolor{black}{Besides, when the strong magnetic field condition is no longer valid, the analytical expressions of Eq.~(\ref{eq:conditional_rotation},~\ref{eq:unconditional_rotation})  are less accurate and  numerical optimization is required.} Finally, the rest of the off-resonance values for $t$ gives unconditional rotations gates about axes on the $x$-$z$ plane that are close to the $z$-axis ($R_Z(\theta)$) with varying rotation angles. 

It is evident that the CPMG sequences allow selective and precise control of nuclear spins as long as the perpendicular component of the hyperfine interaction between the electron and target nuclear spin is nonzero. By setting the CPMG unit time $t$ to be equal to the resonance condition of the  target nuclear spin, Eq.~\eqref{eq:conditional_rotation}, and recursively applying the CPMG unit $N$ times, one can implement a two-qubit gate $CR_X(\phi) $ that conditionally rotates the target nuclear spin  about the $x$-axis by a desired angle $N\phi$. Note that the two-qubit gate $CR_X(\tfrac{\pi}{2})$ is equivalent to a \textsc{cnot} gate up to local operations (see Appendix~\ref{app: A - Effective cnot gate} for further discussion). Similarly, by choosing an off-resonance time $t$ one can apply single-qubit gates to the nuclear spins qubits, where the type of gate (e.g. $R_x(\tilde{\phi})$ or $R_z(\theta)$) is determined by the CPMG unit time $t$, and the angle of rotation depends on the total number $N$ of applied CPMG units. It is \textcolor{black}{worth} noting that, in general, it is preferable that the angle $\phi_k^{\text{CPMG}}$ (or $\tilde{\phi_k}^{\text{CPMG}}$) be either small or close to $2\pi$. The reason is that this allows, by appropriately choosing $N$, to have a total rotation angle $N\phi_k^{\text{CPMG}}$ (or $N\tilde{\phi}_k^{\text{CPMG}}$) that is close to any target rotation angle, which increases the overall gate fidelity $\mathcal{F}$. However, a rotation angle $\phi_k^{\text{CPMG}}$ (or $\tilde{\phi}_k^{\text{CPMG}}$) that is extremely small or extremely close to $2\pi$ would not be as desirable since it would result in a larger gate time $T=Nt$, and thus it would reduce the fidelity of the gate.

Despite the many advantages of CPMG-based spin control, there are also drawbacks. One is that the magnitude of the rotation angle obtained with CPMG, $\phi^{\text{CPMG}}$~\eqref{eq:conditional_rotation}, is not small enough to implement high-fidelity gates. Moreover, this angle does not depend on $k$, and thus higher resonance orders cannot be used or combined in a clever way to get as close as desired to the target rotation and improve the resulting gate fidelity. Another disadvantage is that the use of large magnetic field strengths to improve the entangling gate fidelities would also negatively affect the selectivity in the control of different target nuclear spins with similar hyperfine interaction parameters. In other words, two or more nuclear spins with comparable hyperfine interaction parameters under a high magnetic field would also have similar resonance conditions. In that case, setting the value of the CPMG unit time $t$ to couple the electron spin to the target nuclear spin would also undesirably couple it to the other nuclear spins, hence hindering the spin selectivity. It is in view of these limitations that we explore other types of dynamical decoupling sequences in the following sections.

\section{Controlling nuclear spins with UDD sequences}\label{sec:IV Controlling nuclear spins with UDD sequences}

An Uhrig dynamical decoupling (UDD) sequence~\cite{UhrigPRL2007} is a series of $\pi$ pulses ($\pi$ rotations around the $x$- and/or $y$-axes) which, in contrast to CPMG, are not equidistantly spaced. Instead, their fractional locations are given by
\begin{equation}\label{eq:UDD_fractional_pulse_location_eq}
\delta_{j}=\sin ^{2}(\pi j /(2 n+2)),
\end{equation}
for an $n$ number of pulses (UDD$n$) and unit sequence time $t$ (total time of a single UDD$n$ sequence). Note that for $n=2$ (UDD2) the sequence is exactly equal to the building block of CPMG. Moreover, with each additional pulse, the UDD sequence successively cancels higher orders of a time expansion for any decoherence model~\cite{Yang2008a,Lee2008}. 

\textcolor{black}{In type Ib diamonds, the dominant decoherence source is the spin bath of substitutional nitrogen defects (P1 centers), whose noise spectrum is of Lorentzian shape and decreases very slowly at high frequencies. For such a noise spectrum, previous studies~\cite{Cywinski2008,Uhrig2008,Pasini2010a} have shown that UDD with a large number of pulses tends to perform sub-optimally in comparison to CPMG, and thus simpler sequences (small $n$) are preferable to higher order ones~\cite{Uhrig2008}. For diamonds of the type IIa, which is the type used in the experiments by the Delft group~\cite{vanderSar2012,TaminiauPRL2012,Taminiau:2014aa}, the decoherence is dominated by the hyperfine interaction with the $^{13}$C nuclear spins. Given that in diamond the noise spectrum due to a nuclear spin bath has a hard high-frequency cutoff \cite{ZhaoNanPRB2008,Hall2014}, UDD is expected to perform optimally~\cite{Cywinski2008,Uhrig2008,Pasini2010a}.}

The single unit of a general UDD$n$ can be iterated $N$ times to form a long train of pulses, i.e. $(\mathrm{UDD}n)^N $.
We numerically calculate  the dynamics under UDD applied on the electron spin and find that the nuclear spin evolution satisfies the periodic resonance conditions in a way similar to CPMG. However, its rotation angle and resonance time behave  differently from CPMG. Following an approach similar to the one used in Ref.~\cite{TaminiauPRL2012}, we find analytical expressions for the conditional and unconditional rotation angles and their respective unit times for UDD3 and UDD4 (see Appendix~\ref{app:analytical expressions for UDD}). For UDD3 the resonance unit time and rotation angle are given by
\begin{equation}\label{eq:conditional_rotation_UDD3}
\begin{aligned}
	t_k^{\text{UDD3}} &\approx\frac{2(2k-1)\pi}{2\omega_L-A_{\parallel}}, \\
	\phi_k^{\text{UDD3}} & \approx 2\pi- \frac{2A_{\perp}\left(1-2\cos\left[\frac{(2k-1)\pi}{2\sqrt{2}}\right]\right)}{\omega_L-A_{\parallel}}.
\end{aligned}	
\end{equation}

Similarly to the CPMG case, when $t$ is in the middle of two neighboring resonance values,$\frac{1}{2}(t_k^{\text{UDD3}}+t_{k+1}^{\text{UDD3}})$, the uncoupled nuclear spin rotates unconditionally about the $x$-axis by an angle $\tilde{\phi}^{\text{UDD3}}$. The analytical expressions of these variables are
\begin{equation}\label{eq:unconditional_rotation_UDD3}
\begin{aligned} 
	\tilde{t}_k^{\text{UDD3}} &  \approx\frac{4 k\pi}{2\omega_L-A_{\parallel}},\\
	 \tilde{\phi}_k^{\text{UDD3}} & \approx \frac{2 A_{\perp}}{\left(2 \sqrt{2}+2\right) (\omega_L-A_{\parallel})^2} \left[\pi ^2 A_{\parallel}^2 k^2 \left(\cos \left[\sqrt{2} \pi  k\right]\right.\right.\\
	 &\left.\left. +2 \sqrt{2} \cos \left[\frac{\pi k}{\sqrt{2}}\right]+2\right)+\left(2 \sqrt{2}+3\right) \right.\\
	 &\left.\times A_{\perp}^2 \left(\pi  k+\sin \left[\sqrt{2} \pi  k\right]-2 \sin \left[\frac{\pi  k}{\sqrt{2}}\right]\right)^2 \right]^{1/2}.
\end{aligned}
\end{equation}
Note that for UDD3 and, in general, for any UDD$n$ with odd $n$, the  electron spin does not return back to the initial state after a single unit sequence as required, and thus the number of iterations $N$ of the single unit sequence must be an even number or otherwise the effect on the nuclear spins is naught. Consequently, the analytical expressions for the rotation angles in Eqs. (\ref{eq:conditional_rotation_UDD3}) and (\ref{eq:unconditional_rotation_UDD3}) correspond to a pair of single UDD3 sequences, i.e. (UDD3)$^2$, each with unit sequence time $t_k^{\text{UDD3}}$ (or $\tilde{t}_k^{\text{UDD3}}$).

In contrast to CPMG and UDD3, UDD4 presents two different analytical expressions for the resonance time, each giving different rotation angles. The first set of analytical expressions for the resonance time and rotation angle is
\begin{equation}\label{eq:conditional_rotation_UDD4_1st_set}
\begin{aligned}
t_k^{\text{UDD4}}&\approx \frac{4(2k-1)\pi}{2\omega_L-A_{\parallel}},\\
\phi_k^{\text{UDD4}}&\approx 2\pi-    \frac{2\sqrt{2}A_{\perp}\cos\left[\frac{(2k-1)\sqrt{5}\pi}{4}\right]}{\omega_L -A_{\parallel}},
\end{aligned}
\end{equation}
where the resonance time coincides with the CPMG one  and  the magnitude of the rotation angle at any order $k$ is much smaller than those generated by the CPMG sequence, Eq.~\eqref{eq:conditional_rotation}. On the other hand, the second set of analytical expressions, which does not coincide with the CPMG resonance time, is
\begin{equation}\label{eq:conditional_rotation_UDD4_2nd_set}
\begin{aligned}
\hat{t}_k^{\text{UDD4}}&  \approx\frac{8(2k-1)\pi}{2\omega_L-A_{\parallel}},\\
\hat{\phi}_k^{\text{UDD4}}&\approx    \frac{4 A_{\perp}\cos\left[\frac{(2k-1)\sqrt{5}\pi}{2}\right]}{\omega_L -A_{\parallel}},
\end{aligned}    
\end{equation}
where the angle of rotations $\hat{\phi}_k^{\text{UDD4}}$ are larger than $\phi_k^{\text{UDD4}}$. In fact, with the resonance time $\hat{t}_k^{\text{UDD4}}$ UDD4 performs almost on a par with CPMG in both gate fidelity and total sequence time. Another difference between UDD4 and CPMG (and UDD3 too) is that unconditional rotations about the $x$-axis do not occur every time $t$ is in the middle between any two sequential resonance times $t_k^{\text{UDD4}}$ (or $\hat{t}_k^{\text{UDD4}}$), they only happen at certain times given by
\begin{equation}\label{eq:unconditional_rotation_UDD4}
\begin{aligned}
\tilde{t}_k^{\text{UDD4}}&\approx\frac{16 k\pi}{2\omega_L-A_{\parallel}},\\
\tilde{\phi}_k^{\text{UDD4}} & \approx \frac{2 k A_{\perp}\pi\sqrt{A_{\perp}^2+A_{\parallel}^2\cos \left[ k\sqrt{5}\pi \right]^2 }}{(\omega_L-A_{\parallel}^2)}.
\end{aligned}
\end{equation}
In general, for both CPMG and any UDD\textit{n}, the sequence unit times $\tilde{t}_k$ that generate unconditional rotations about the $x$-axis are more sensitive to timing imprecision. As discussed before, the timing sensitivity is connected to the sharpness of the dips and peaks of the rotation axes $x$ projection as shown in Fig.~\ref{fig:geometric_representation} (see also Appendix~\ref{app:analytical expressions for UDD}). As a result, the analytical expressions for the sequence unit time that generates unconditional rotations, and the corresponding rotation angles, are less precise approximations in comparison to the analytical expressions for the conditional rotations, and thus they should be used as initial inputs of a numerical optimization algorithm that would give more exacts values.

\begin{figure*} [t]
\includegraphics[width=\linewidth]{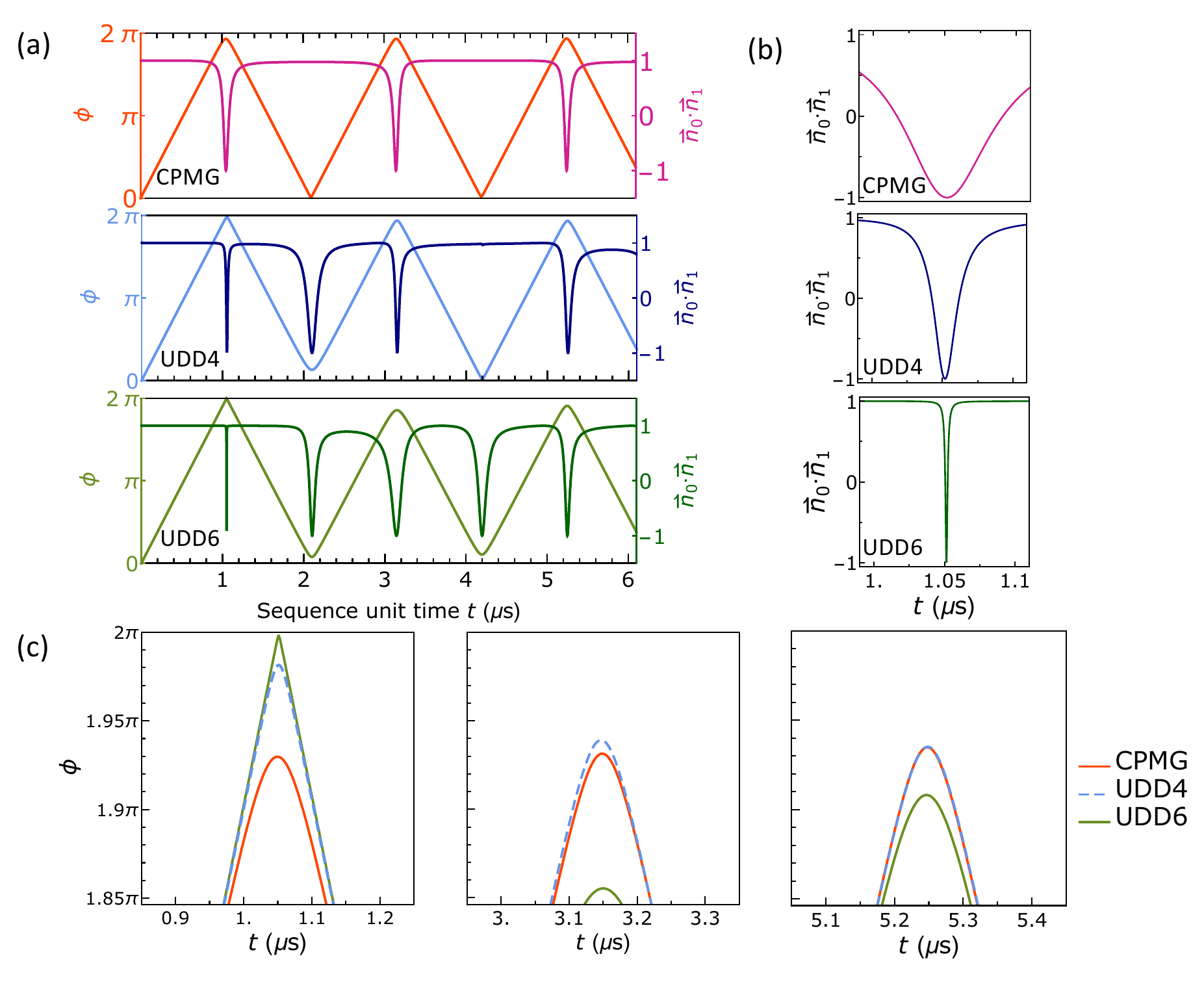} 
\caption{\textcolor{black}{(a) The dot product $\vec{n}_0\cdot\vec{n}_1$ of rotation axes and the rotation angle $\phi$ of a target nuclear spin under the CPMG (top), UDD4 (middle), UDD6 (bottom) sequences, against the unit sequence time. As in Fig.~\ref{fig:geometric_representation}, the periodic dips of $\vec{n}_0\cdot\vec{n}_1$ indicate conditional rotations and the flat regions indicate unconditional ones. For the numerical simulation we set $A_{\parallel}=A_{\perp}=0.1\omega_L, \text{ where }\omega_L/2\pi= 1\text{MHz}$. 
(b) Close-up of  $\vec{n}_0\cdot\vec{n}_1$ around the first resonant time in (a). Clearly, at the first resonant time, UDD has narrower spectral widths than CPMG.
(c) Close-up of rotation angles of CPMG (blue), UDD4 (red dashed) and UDD6 (green) at the first three resonant times ($t_1,t_2,t_3$). The resonant rotation angles ($\phi_i$) of CPMG are relatively constant, while those of UDD are varying. The resonant rotation angles of CPMG at $t_1,t_2,t_3$ are $(\phi_1,\phi_2,\phi_3)=(1.93\pi,1.93\pi,1.94\pi)$, of UDD4 are $(\phi_1,\phi_2,\phi_3)=(1.98\pi,1.94\pi,1.94\pi)$ and of UDD6 are $(\phi_1,\phi_2,\phi_3)=(1.996\pi,1.78\pi,1.91\pi)$.}}
\label{fig:geo_rep_UDD_DD}
\end{figure*} 

We do not provide analytical expressions for the resonance times (or corresponding rotation angles) for UDD\textit{n} with $n\geq 5$ due to the increased complexity in the expressions. Instead, the numerical comparison between UDD$n$ and CPMG is in Fig.~\ref{fig:geo_rep_UDD_DD}, where the rotation angles and resonance times of the target nuclear spin are shown for both cases. Again, since the general UDD is not equidistantly spaced, we use the basic sequence time $t$ as the unit time instead of the variable inter-pulse time $\tau$. The first resonance time $t_1$ (first dips of  \textcolor{black}{ $\vec{n}_0 \cdot\vec{n}_1$} in Fig.~\ref{fig:geo_rep_UDD_DD}(a)) for CPMG, UDD4, and UDD6 is the same.
On the other hand, as shown in Figs.~\ref{fig:geo_rep_UDD_DD}(b,c), the rotation angle $\phi$ and spin selectivity (full width at half minimum of the curve for the dot product of the rotation axes, $\vec{n}_0 \cdot \vec{n}_1$) vary for different UDD$n$ sequences.
\textcolor{black}{ Fig. ~\ref{fig:geo_rep_UDD_DD}(c) shows the rotation angles around the first three orders of the resonant time $t_k$ with $k=1,2,3$ for CPMG, UDD4 and UDD6. It is evident that  the rotation angles  at  $t_k$ are almost constant for CPMG yet varying for UDD4 and UDD6. The closer the magnitudes of $\phi_k$ is to $2\pi$ the smaller the effective rotation angle is.}
Therefore, we can choose a UDD sequence with a rotation angle close enough to $2\pi$ to implement a total gate with higher fidelity.
In order to have overall short gate times, hereafter we will use the first resonance time $t_1$ to implement the coupling gate for nuclear spins under both CPMG and UDD sequences, unless stated otherwise. Moreover, we will only consider UDD$n$ sequences with $n\leq 6$ in order to avoid unnecessarily long sequences and also to maintain the efficacy of the sequence in protecting the electron spin from noise with soft high-frequency cutoff~\cite{Uhrig2008}.

\subsection{Nuclear spin selectivity enhancement using UDD}\label{sec:IV-B Nuclear spin selectivity enhancement using UDD}

\begin{figure*} [tb]
\includegraphics[width=\linewidth]{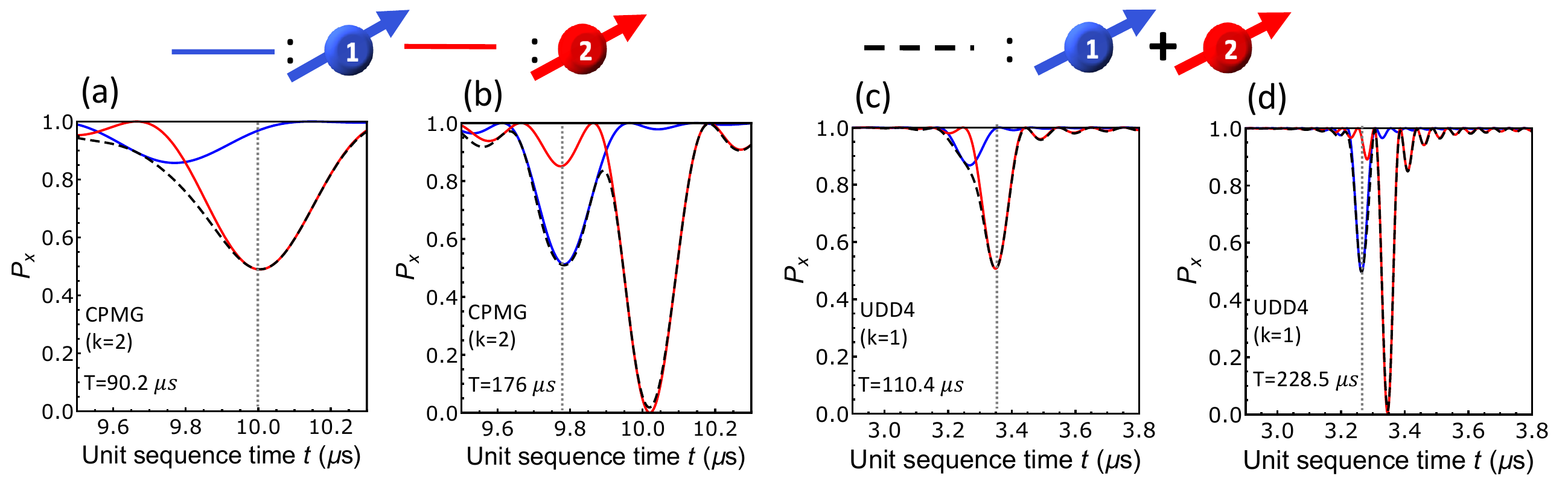} 
\caption{\textcolor{black}{Probability $P_x$ of preserving the initial $\ket{x}$ state of the electron spin when it interacts with one nuclear spin (blue and red curves) and with two nuclear spins simultaneously (black dashed curve) after (a,b) CPMG and (c,d) UDD4 pulse sequences. The CPMG (UDD4) sequence is formed by $N_{\text{CPMG}}$ ($N_{\text{UDD}}$) copies of its basic unit sequence. The hyperfine interaction parameters $({A_{\parallel}}/{2\pi},{A_{\perp}}/{2\pi})$ of the two nuclear spins interacting with the electron spin are extracted from Ref.~\cite{TaminiauPRL2012} and are: (15.3, 12.9) kHz for spin-1 (blue curve), and (30.6, 25.7) kHz for spin-2 (red curve). The calculation assumes a relatively strong external magnetic field, $\omega_L/2\pi=314$ kHz. Each panel shows the type of pulse sequence used in the numerical simulation and the total sequence time $T$.
In each panel, when the unit sequence time $t$ is chosen as indicated by the dotted vertical line, the targeted spin (spin-1 in (b,d) and spin-2 in (a,c)) is maximally entangled with the electron spin ($P_x\rightarrow1/2$). The intersection of the curve of untargeted spin (spin-1 in (a,c) and spin-2 in (b,d)) with the dotted vertical line represents how much the untargeted spin is decoupled. $P_x=1$ for the untargeted spin represents complete decoupling of the untargeted spin. (a,b) The CPMG sequence fails to fully control each nuclear spin individually without affecting the other one, due to the small $P_x$ ($P_x=$0.8 in (b)) of the untargeted spin. Here we use a higher resonance order $k=2$ (corresponding to larger unit sequence time $t_k$), Eq.~\eqref{eq:conditional_rotation}, which improves the selectivity of the CPMG sequence. (c,d) Using the UDD4 sequence on the electron spin increases the coupling selectivity with each nuclear spin. The number of iterations, $N_i^{(j)}$ with $i\in \{\text{CPMG,UDD}\}$ and $j$ the panel label, are chosen such that the electron spin is maximally entangled with one of the nuclear spins. The iteration numbers used in each panel are $N_{\text{CPMG}}^{(a)}=9$, $N_{\text{CPMG}}^{(b)}=18$, $N_{\text{UDD}}^{(c)}=33$, and $N_{\text{UDD}}^{(d)}=70$. }
}
\label{fig:selectivity_2spins}
\end{figure*} 

A good nuclear spin selectivity, in the context of the techniques discussed in this work, implies the successful coupling of the electron spin with a target nuclear spin and the simultaneous decoupling from the rest of the nuclear spin bath. However, when two or more nuclear spins surrounding the central electron spin have similar hyperfine parameter values, it becomes challenging to couple the electron spin to one of those nuclear spins without coupling to the other. 
In the case of CPMG, the spin selectivity can be improved by using a higher resonance order $k$~\cite{TaminiauPRL2012}, Eq.~\eqref{eq:conditional_rotation}, i.e. larger unit sequence time $t_k$. However, a larger minimum pulse interval implies a reduction in the `dynamical decoupling limit'~\cite{Khodjasteh2011,Biercuk2011}, which is the highest-frequency component of the noise power spectral density that can be successfully suppressed by dynamical decoupling. As a result, a CPMG sequence with larger $t_k$ will unavoidably underperform (see Sec.~\ref{Sec:decoupling_power_UDD_CPMG} for further discussion).
Alternatively, given that the spin selectivity (full width at half maximum of the curve for $\vec{n}_0\cdot\vec{n}_1$ of the target nuclear spin, see Fig.~\ref{fig:geo_rep_UDD_DD}(b)), varies for different UDD$n$ sequences, UDD-based control is versatile enough to individually control a target nuclear spin while decoupling the electron spin from the rest of the spin bath, without the use of higher resonance orders.

As an example of the finer spin selectivity of UDD, we simulate the interaction of two $^{13}$C nuclear spins with the central electron spin of an NV center under dynamical decoupling sequences \textcolor{black}{(see Appendix \ref{appendix:selectivity_3_spin} for the case of three nuclear spins coupled to the central electron spin)}. We calculate the coupling-decoupling rate of the electron spin using the system's coherence function~\cite{ZhaoNanPRB2008} $L(t)=\operatorname{Tr}\left[\rho(t)S^{+}\right]/\operatorname{Tr}\left[\rho(0) S^{+}\right]$, where $S_+ = S_x + iS_y$ is a spin ladder operator, and $\rho(t)$ is the density matrix of the system comprising the electron spin and two nuclear spins at time $t$. For the numerical calculations we set the nuclear Larmor frequency equal to $\omega_L/2\pi$=314 kHz, and we take the hyperfine interaction parameters $({A_{\parallel}}/{2\pi},{A_{\perp}}{2\pi})$ from Ref.~\cite{TaminiauPRL2012}: (15.3, 12,9) kHz and (30.6, 25.7) kHz for the first and second nuclear spins, respectively.
Assuming an inter-nuclear distance such that the nuclear-nuclear interaction is much weaker than the electron-nuclear interaction, we neglect the former in our calculations. Consequently, if we assume that the system is initialized in a product state with the electron state being $\ket{x}=(\ket{m_s=0}+\ket{m_s=-1})/\sqrt{2}$, then the probability $P_x$ of preserving the initial electron state at time $t$ is given by $P_x=(1+L(t))/2$. In Fig.~\ref{fig:selectivity_2spins} we plot the probability $P_x$ after CPMG (Figs.~\ref{fig:selectivity_2spins}(a,b)) and UDD4 (Figs.~\ref{fig:selectivity_2spins}(c,d)) sequences with $N$ iterations of their respective basic unit sequences.
Evidently, a probability $P_x$ equal to 1 indicates that the electron is decoupled from the nuclear spins, and that is true for most values of the unit sequence time $t$ in Fig.~\ref{fig:selectivity_2spins}. However, for certain values of $t$ the sequence is in resonance with one of the nuclear spins, which corresponds to a sharp dip in $P_x$ as shown in Fig.~\ref{fig:selectivity_2spins}. At those resonance values of $t$, Eqs.~(\ref{eq:conditional_rotation},\ref{eq:conditional_rotation_UDD4_1st_set}), the rotation axes $\vec{n}_0$ and $\vec{n}_1$ for the target nuclear spin are approximately antiparallel ($\vec{n}_0\cdot\vec{n}_1=-1$,which corresponds to $P_x\approx 0.5$), and thus the resulting conditional rotation entangles the target nuclear spin with the electron spin.  In order to improve the spin selectivity of CPMG we use a higher resonance order $k=2$. Moreover, the number of iterations, $N_i$ with $i\in \{\text{CPMG,UDD}\}$, are chosen such that the electron spin is maximally entangled with one of the nuclear spins ($P_x\rightarrow 0.5$), see Fig.~\ref{fig:selectivity_2spins}.
Note that, in contrast to the CPMG case (Figs.~\ref{fig:selectivity_2spins}(a,b)), when the signal of the target nuclear spin, which is at resonance with the UDD4 sequence (Figs.\ref{fig:selectivity_2spins}(c,d)), is near $P_x=0.5$, the signal of the nuclear spin is effectively at $P_x=1$. This means that the nuclear spins can be individually controlled without significantly affecting each other in the process, \textcolor{black}{thus minimizing the crosstalk error between spin qubits, which is a necessary condition for realizing the control of a multi-nuclear spin register~\cite{BradleyPRX2019}}. Moreover, the total sequence time $T$ of UDD4 is only slightly larger than that of CPMG, making it overall more appealing.

\subsection{Decoupling power of UDD versus CPMG in the spin selectivity enhancing case}\label{Sec:decoupling_power_UDD_CPMG}

It has been shown in the literature that in the presence of noise with a soft high-frequency cutoff CPMG outperforms UDD~\cite{Cywinski2008,Uhrig2008,Biercuk2011,Wang2012a,Pasini2010a}. However, as mentioned in the previous section, in the particular case of spin selectivity enhancement it is necessary to use a higher resonance order $k$ for CPMG (larger interpulse period). Higher values of $k$, in addition to making the sequences longer, affect negatively their decoupling performance. This becomes evident when we consider the electron spin decoherence under pulse sequences. Accordingly, we quantify the electron spin (qubit) coherence following the formulation for measuring coherence under a dephasing Hamiltonian introduced in Refs.~\cite{UhrigPRL2007,Uhrig2008,Cywinski2008}. In general, an initial qubit state along the $x$-axis of the Bloch sphere accumulates a random phase due to its interaction with the environment. The coherence of the state after a time $T$ is given by $\overline{\vert\ L(t) \vert}=e^{-\chi(T)}$, where $\overline{\vert \ldots \vert}$ is the ensemble average and $L(t)$ is the previously defined coherence function. As shown in Refs.~\cite{UhrigPRL2007,Uhrig2008,Cywinski2008}, the function in the exponent of the coherence function is $\chi(T)=\tfrac{2}{\pi}\int_0^{\infty}\tfrac{S(\omega)}{\omega^2}F(\omega T)\mathrm{d}\omega$, where $S(\omega)$ is the power spectral density of the noise, and $F(\omega T)$ is known as the `filter function' and describes the influence of the pulse sequence on the qubit decoherence. Therefore, to characterize the coherence-preserving power of any pulse sequence with total time $T$, it suffices to calculate its filter function $F(\omega T)$.

For a general sequence of $n$ $\pi$ pulses which are applied at the instants of time $\delta_j T$ with $j\in\{1,2,\ldots ,n\}$, so that the total sequence time $T$ is divided into $n+1$ subintervals, the filter function is~\cite{UhrigPRL2007,Uhrig2008}
\begin{equation}\label{eq:filter_function_general}
    F(\omega T)=\left | 1+(-1)^{n+1}e^{i \omega T}+2\sum_{j=1}^{n}(-1)^j e^{i\delta_j\omega T}\right | ^2.
\end{equation}
Here we are assuming instantaneous pulses, which is a good approximation as long as the duration of each pulse is smaller than the smallest interval between pulses~\cite{Cywinski2008}. This is the case for NV centers, where $\pi$-pulses can be implemented in less than 10 ns and with a fidelity above 99\%~\cite{deLange60,Dobrovitski2010a}. \textcolor{black}{Moreover, for the numerical simulations we use noiseless pulses and assume that there are no other noise sources acting on the system}. Now, for an $n$-pulse CPMG sequence the fractional pulse locations are $\delta_j=(j-1/2)/n$. However, given that in this work we consider the number of iterations $N$ of a basic sequence unit $(\tau-\pi-2\tau-\pi-\tau)$ instead of the total number of pulses, the fractional pulse locations for $N$ iterations of the basic CPMG unit (CPMG$^N$) would be $\delta_j=(j-1/2)/(2N)$. Therefore, after some algebra, the filter function for a CPMG$^N$ sequence is
\begin{equation}\label{eq:filter_function_CPMG^N}
    F^{\mathrm{CPMG}}_{N}(\omega T)=16\sec^2\left(\frac{\omega T}{4N}\right)\sin^2\left(\frac{\omega T}{2}\right)\sin^4\left(\frac{\omega T}{8N} \right).
\end{equation}

Alternatively, for a UDD sequence with $n$ pulses (UDD$n$) the fractional pulse locations are $\delta_j=\sin^2[\pi j/(2n+2)]$. But then again, in this work we consider $N$ iterations of a basic sequence unit UDD$n$, and thus the fractional pulse locations are given by $\delta_{l n+j}=l/N+\sin^2[\pi j/(2n+2)]/N$. The filter function for a (UDD$n$)$^N$ sequence then is
\begin{equation}\label{eq:filter_function_UDDn^N}
\begin{aligned}
    F_N^{\text{UDD\textit{n}}}(\omega T)=\vert & 1+(-1)^{n N+1}e^{i \omega T}\\
    & +2\sum_{l=0}^{N-1}\sum_{j=1}^{n}(-1)^{l n +j}e^{i \delta_{l n+j}\omega T}\vert^2.
    \end{aligned}
\end{equation}

\begin{figure} [tb]
\includegraphics[width=\linewidth]{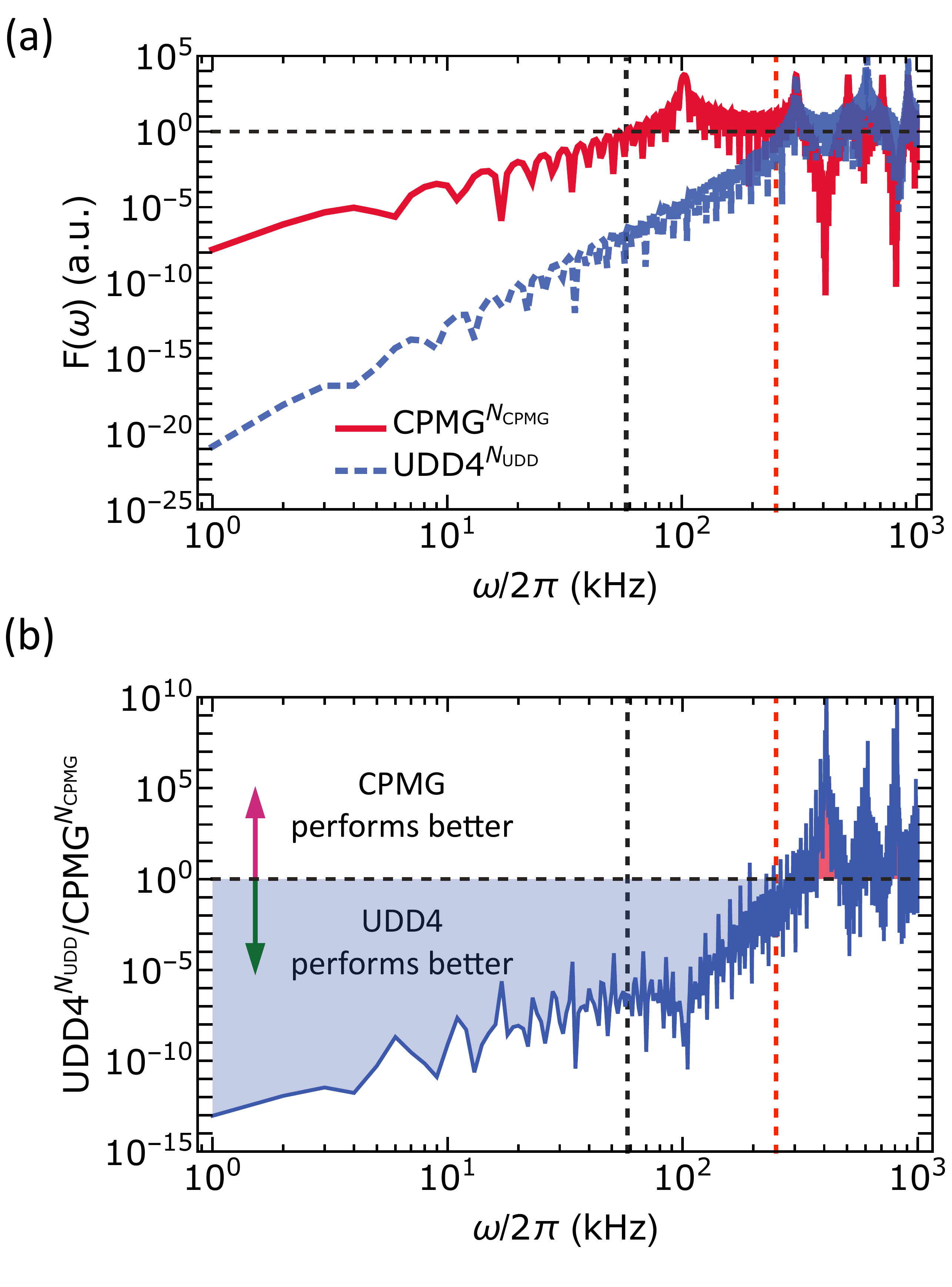} 
\caption{Comparison of the CPMG and UDD4 filter functions plotted against the noise frequency $\omega/2\pi$. The pulse sequences are the same ones used in Fig.~\ref{fig:selectivity_2spins}, where the total time of the CPMG (UDD4) sequence is 90.2 $\mu$s (110.4 $\mu$s) and the number of iterations is $N_{\text{CPMG}}=9$ ($N_{\text{UDD}}=33$). (a) Numerically calculated filter functions for CPMG and UDD4. The horizontal dashed black line indicates $F(\omega)=1$. (b) The quotient between the numerical values of UDD4 and CPMG. A quotient equal to 1 indicates equal filter functions, a quotient less than 1 (light blue shading) corresponds to UDD4 outperforming CPMG and vice versa for a quotient greater than 1 (magenta shading). In the frequency interval $\sim$[60,250] kHz (between the vertical black and red dashed lines) the CPMG filter function is, on average, equal or greater than 1, thus losing its error-suppressing capability. In that same frequency interval UDD4 clearly outperforms CPMG. For higher frequencies (to the right of the red dashed line $\sim$250 kHz) CPMG outperforms UDD4 for certain sporadic intervals, however, the noise spectral weight in such intervals is comparatively much smaller.}
\label{fig:FilterFunctionUDD}
\end{figure}

With the above expressions we proceed to compare the filter functions of the pulse sequences used in Figs.~\ref{fig:selectivity_2spins}(a,c), which is shown in Fig.~\ref{fig:FilterFunctionUDD} (the filter functions of the pulse sequences used in Figs.~\ref{fig:selectivity_2spins}(b,d) gave similar results). The vertical black and red dashed lines in Figs.~\ref{fig:FilterFunctionUDD}(a,b) mark the frequency interval $\sim$[60,250] kHz where the CPMG filter function is greater than or equal to 1, i.e. CPMG fails to decouple the electron spin from the spin bath. In that same region, with non-negligible noise spectral weight, UDD4 clearly outperforms CPMG. This shows that, in this type of scenario, UDD4 not only provides better spin selectivity, but also better noise-suppression compared to CPMG.

\subsection{Robustness under pulse timing errors in strong magnetic fields}\label{sec: IV-C Nuclear spin control under very strong magnetic field}

There are scenarios where the use of very strong magnetic fields is advantageous, e.g. in order to suppress undesired transverse couplings. In those cases, the use of dynamical decoupling sequences to conditionally control nuclear spins becomes more sensitive to pulse timing errors due to shorter time intervals between pulses that can get close to the hardware temporal resolution limit. 
However, considering that whenever the inner product of the target nuclear spin rotation axes ($\vec{n}_0\cdot\vec{n}_1$) is equal to -1, the nuclear spin rotation is conditional, then it is possible to make the sequence more resistant to pulse timing error by requiring that the gradient around the point where $\vec{n}_0\cdot\vec{n}_1=-1$ be as small as possible. 
In other words, since the spin selectivity depends on the full width at half minimum of the aforementioned curve, we slightly relinquish the spin selectivity in order to obtain a sequence that is more resistant to pulse timing error. 

\begin{figure} [tb]
\includegraphics[width=0.8\linewidth]{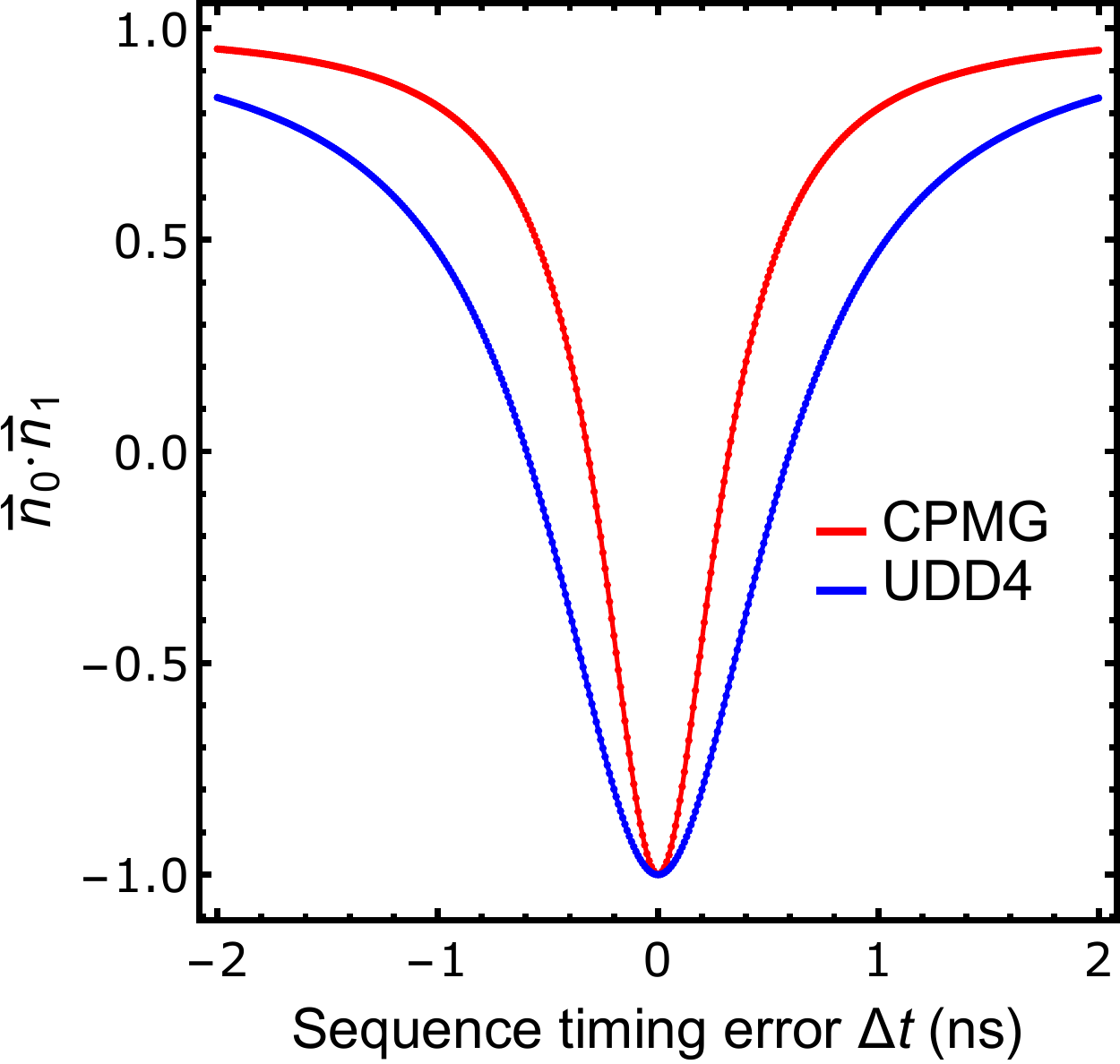} 
\caption{Dot product of the nuclear spin rotation axes, $\vec{n}_0 \cdot \vec{n}_1$, as a function of timing error of unit sequence, centered at the resonance times $t_1^{\text{CPMG}}$ and  $\hat{t}_1^{\text{UDD}4}$  of CPMG (blue curve) and UDD4 (red curve)  sequences, respectively. In the numerical calculations we use a strong external magnetic field (corresponding to a nuclear Larmor frequency of $\omega_L/2\pi=5$MHz) and hyperfine parameters $A_{\parallel}/2\pi=A_{\perp}/2\pi=25~\text{kHz}$. The wider dip given by the UDD4 sequence allows for somewhat larger degree of error in the pulse timing.}
\label{fig:ultra_high_field}
\end{figure}

To illustrate the previous point, Fig.~\ref{fig:ultra_high_field} shows the inner product of the rotation axes, $\vec{n}_0\cdot\vec{n}_1$, of a nuclear spin being controlled via CPMG and UDD4 sequences under a strong magnetic field (corresponding to a nuclear Larmor frequency of $\omega_L/2\pi=5$ MHz). For UDD4 we use the first order resonance time $\hat{t}_1^{\text{UDD}4}$ of the second set (see Eq. (\ref{eq:conditional_rotation_UDD4_2nd_set})) and for CPMG we use its first order resonance time $t_1^{\text{CPMG}}$. Evidently, the CPMG sequence produces a sharp deep, which doesn't change regardless of the order of the chosen resonance time, whereas UDD4 gives a wider dip that corresponds to a control sequence less sensitive to pulse timing error. Note that $\hat{t}_1^{\text{UDD}4}=2t_1^{\text{CPMG}}$, meaning UDD4 requires longer time  to guarantee its superior robustness against timing error.  

\section{{Hybrid} sequences: CPMG+UDD for high fidelity gates and wider spin control range}\label{sec:V Controlling nuclear spins with hybrid sequences}

\begin{figure*} [t]
\includegraphics[width=0.82\linewidth]{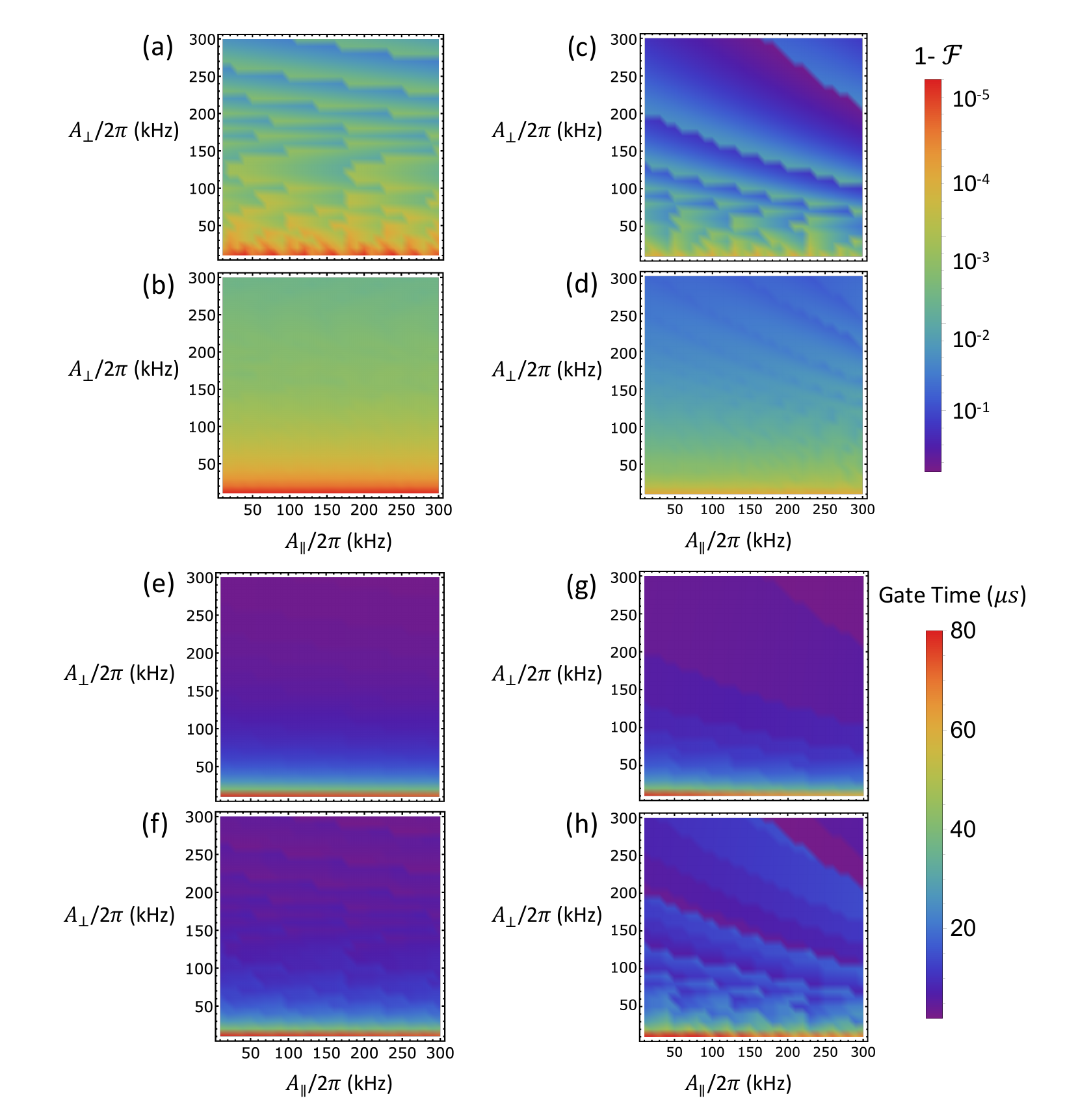} 
\caption{\textcolor{black}{(a-d) Infidelity ($1-\mathcal{F}$) and  (e-h) gate time of  $CR_X(\tfrac{\pi}{2})$ gate, where (a,c,e,g) were obtained with the CPMG sequence and (b,d,f,h) with the CPMG+UDD4 hybrid sequences. The Larmor frequencies used are  (a,b,e,f) $\omega_L/2\pi=2.0$ MHz (relatively strong  magnetic field) and (c,d,g,h) $\omega_L/2\pi=0.5$ MHz (relatively weak magnetic field).
The axes correspond to the the parallel and perpendicular components of the  hyperfine interaction strength, $A_{\parallel}/2\pi$ ($x$-axis) and $A_{\perp}/2\pi$ ($y$-axis), in the range of 10 kHz to 300 kHz.
For each point of the contour plots we have calculated the necessary number of iterations $N_i$ ($i\in \{\mathrm{CPMG},\mathrm{UDD4}\}$) for each type of sequence such that the resulting gate is a $CR_X(\tfrac{\pi}{2})$ with the shortest gate time that achieves maximal fidelity for the given set of parameters (see Appendix \ref{appendix:N_values} for the values of $N_i$). (a,c) The CPMG-based $CR_X(\tfrac{\pi}{2})$ gate infidelity is considerably increased when the magnetic field is weak, especially for nuclear spins with stronger hyperfine parameters. (b,d) The CPMG+UDD4-based $CR_X(\tfrac{\pi}{2})$ gate infidelity is less affected by the lower magnetic field due to the smaller rotation angle of UDD4. Under the same magnetic field, the hybrid sequence CPMG+UDD4 allows more  robust control  in a broader hyperfine coupling parameter range.  (e,g) CPMG-based and (f,h) CPMG+UDD4-based $CR_X(\tfrac{\pi}{2})$ gate times. The hybrid sequence CPMG+UDD4 is always slightly longer than the CPMG sequence alone.}
}
\label{fig:2D_fidelity_time_CNOT}
\end{figure*} 

Given that CPMG offers fast yet non ideal large angle rotations and UDD offers slow but desirable small angle rotations, the combination of both CPMG and UDD sequences is an attractive solution to construct both fast and high fidelity gates. 
We refer to  such combinations of CPMG and UDD as \textit{hybrid} sequences. These are based on several iterations of basic CPMG units to form rotations close to the desired gate, followed by few iterations of single basic UDD units to get as close as possible to the target gate. The resulting rotation angle $\Theta$ is given by
\begin{equation}
	\Theta = N_{\text{CPMG}} \theta_{\text{CPMG}}+N_{\text{UDD}} \theta_{\text{UDD}},
\end{equation}
where $\theta_{\text{CPMG(UDD)}}$ and $N_{\text{CPMG(UDD)}}$ are the rotation angle and integer number of iterations of the CPMG (UDD) sequence, respectively. 
The recipe for choosing $N_{\text{CPMG}}$ and $N_{\text{UDD4}}$ is to start with the value for $N_{\text{CPMG}}$ that makes the resulting gate as close to the target gate as possible. Then we perform a simple numerical optimization where we perturb the value for $N_{\text{CPMG}}$ previously found and add a variable number of iterations of the UDD4 sequences (constrained to $N_{\text{UDD4}}\leq 6$) in such a way that the resulting gate fidelity is maximum.
These hybrid  sequences take advantage of the large rotation speed of CPMG and the small and more precise rotation angles of UDD, giving an overall fast and high-fidelity two-qubit entangling gate. 

Fig.~\ref{fig:2D_fidelity_time_CNOT} shows the gate infidelity and gate time for a two-qubit $CR_X(\tfrac{\pi}{2})$ obtained with the CPMG sequence and with the hybrid CPMG+UDD4 sequence. The gate infidelity is defined as~\cite{PEDERSEN200747} $1-\mathcal{F}=1-\tfrac{1}{n(n+1)}[\Tr(U^{\dagger}U)+\vert\Tr(U_0^{\dagger}U)\vert^2]$, where $n$ is the Hilbert space dimension, $U$ is the generated gate, and $U_0$ is the desired gate. The infidelity and gate time are sampled on a range of hyperfine parameter values and for relatively weak and strong magnetic field strengths. The CPMG-based $CR_X(\tfrac{\pi}{2})$ gate has lower fidelity when the external magnetic field is relatively weak (Fig.~\ref{fig:2D_fidelity_time_CNOT}(c)). In contrast, the CPMG+UDD4-based $CR_X(\tfrac{\pi}{2})$ gives relatively high gate fidelity under weak magnetic field strength, above 99\% as shown in Fig.~\ref{fig:2D_fidelity_time_CNOT}(d). \textcolor{black}{Moreover, as shown in Fig.~\ref{fig:2D_fidelity_time_CNOT}(a,c) the large effective rotation angle incurs the granularity pattern of the infidelity.  On the other hand, due to  the smaller effective rotation angle of UDD4, the high fidelity of the hybrid CPMG+UDD4 sequence persists for a broad hyperfine parameter range and the landscape is smoother, as shown in Fig.~\ref{fig:2D_fidelity_time_CNOT}(b,d).} Figs.~\ref{fig:2D_fidelity_time_CNOT}(e-h) show the gate times for $CR_X(\tfrac{\pi}{2})$ obtained with both type of sequences. The CPMG+UDD4-based $CR_X(\tfrac{\pi}{2})$ has only a slightly longer gate time than the CPMG-based one, confirming that the combination of CPMG and UDD sequences gives an overall fast and high-fidelity $CR_X(\tfrac{\pi}{2})$ gate.

\begin{figure} [tb]
\includegraphics[width=\linewidth]{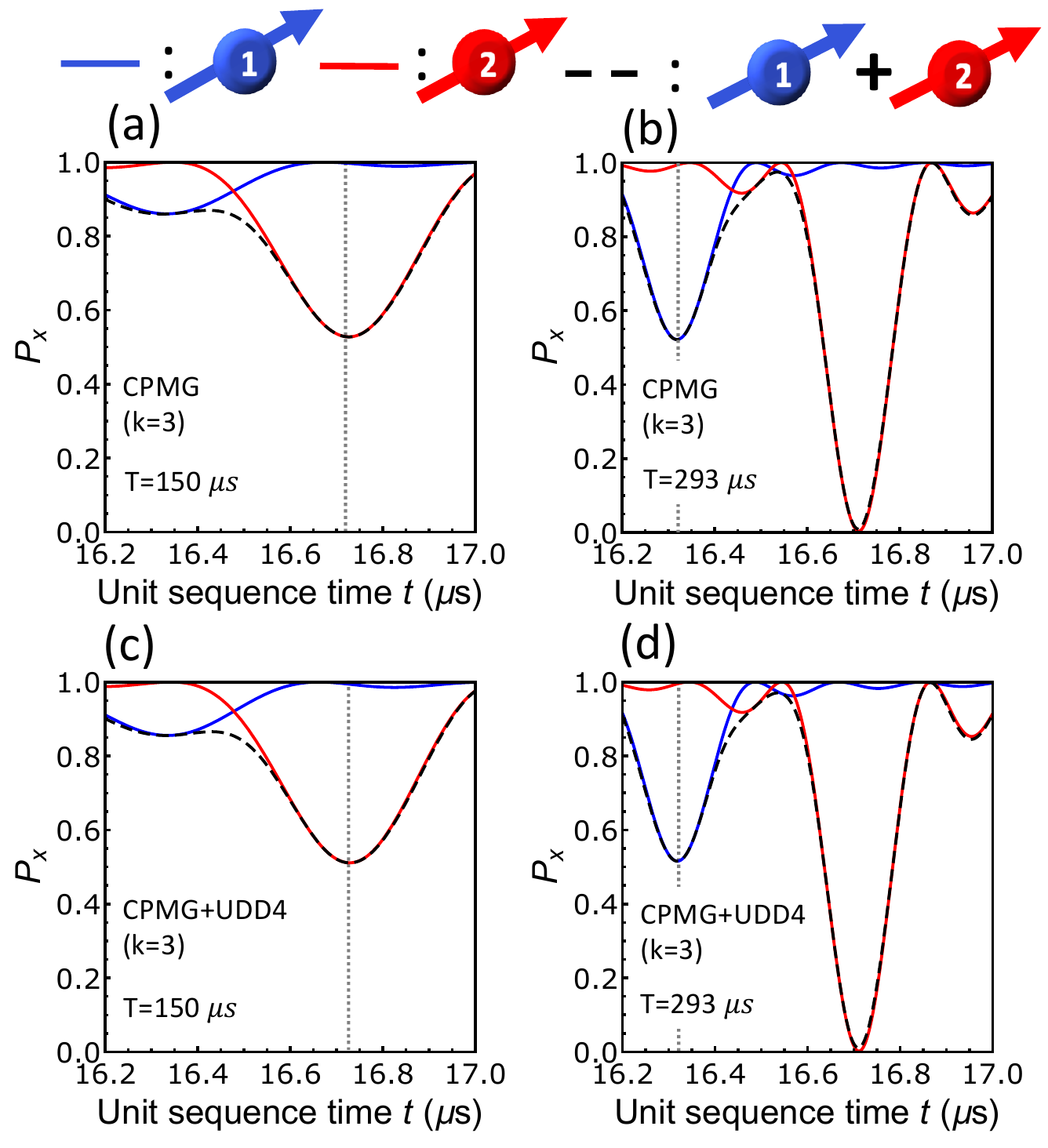} 
\caption{Probability $P_x$ of preserving initial $\ket{x}$ state of the electron spin after the (a,b) CPMG protocol and the (c,d) CPMG+UDD4 hybrid protocol. The $k=3$ (third resonance) is used for both CPMG and UDD4. The hyperfine parameters of nuclear spin 1 (blue) and nuclear spin 2 (red curve) and magnetic field setup are the same as in Fig.~\ref{fig:selectivity_2spins}. Simulation in the simultaneous presence of two nuclear spins is plotted in black dashed curve. The choice of third order resonance here results in well separated resonance time between the two nuclear spins, thus improving the spin selectivity. \textcolor{black}{For each plot we use} (a) $N_{\text{CPMG}}=9$, (c) $N_{\text{CPMG}}=8,N_{\text{UDD4}}=1$ (b) $N_{\text{CPMG}}=18$, (d) $N_{\text{CPMG}}=17, N_{\text{UDD4}}=1$. The CPMG+UDD4 hybrid protocol  shows similar spin selectivity and equivalent gate time compared to the pure CPMG protocol. \textcolor{black}{The lack of selectivity enhancement of the hybrid protocol is due to its predominant CPMG composition.}}
\label{fig:selectivity_hybrid}
\end{figure}

\begin{figure}   
\includegraphics[width=\linewidth]{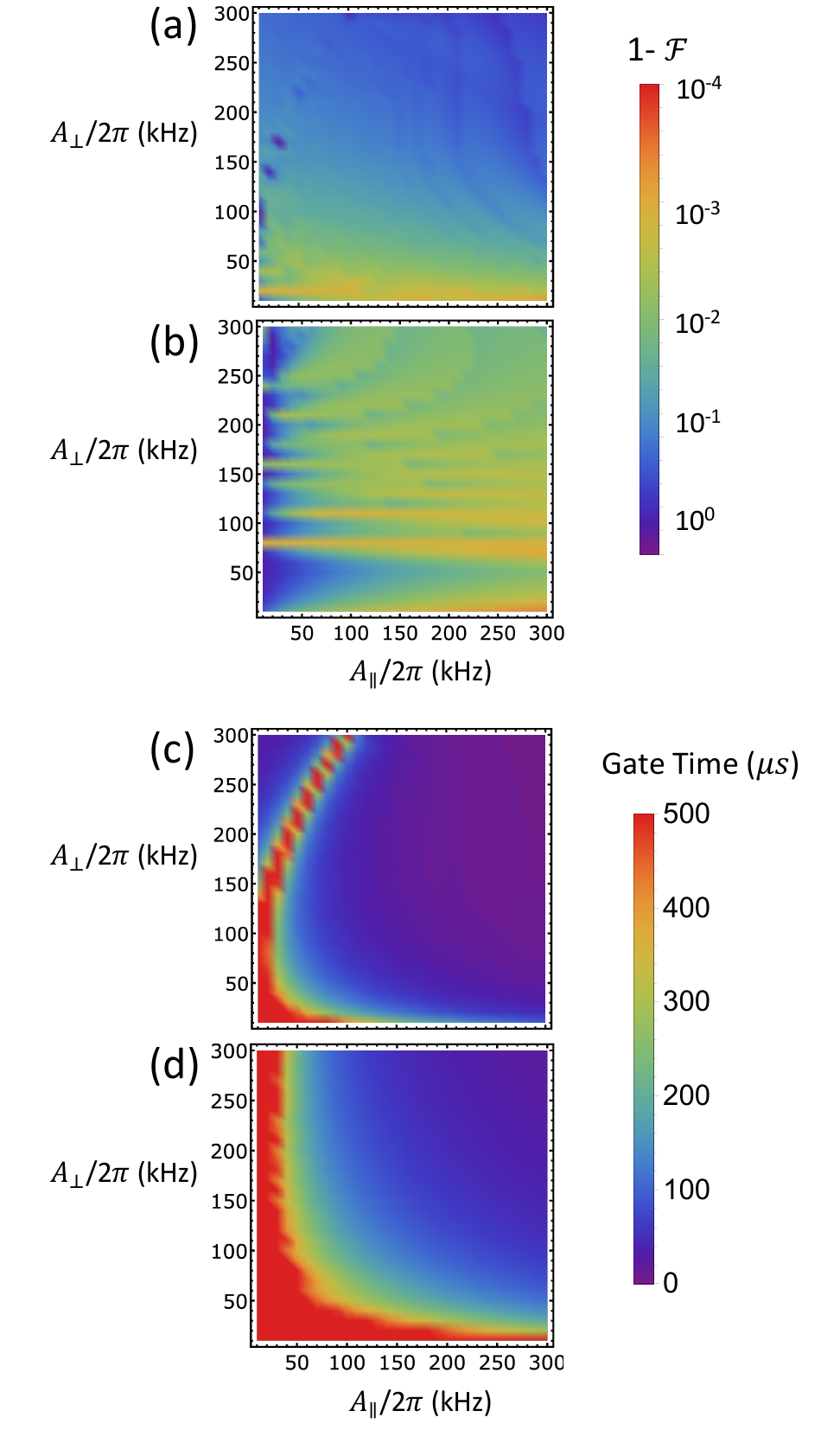} 
\caption{\textcolor{black}{(a,b) Infidelity ($1-\mathcal{F}$)  and (c,d) gate time  of single-qubit $R_X(\frac{\pi}{2})$ gate under (a,c) $\omega_L/2\pi=0.5$ MHz (relatively weak magnetic field) and (b,d) $\omega_L/2\pi=2.0$ MHz (relatively strong magnetic field). The nuclear spin single-qubit gate is implemented via the unconditional rotation caused by the CPMG sequence on the electron spin. The axes correspond to the parallel and perpendicular components of the hyperfine interaction strength $A_{\parallel}/2\pi$ (x-axis) and $A_{\perp}/2\pi$ (y-axis). Decreasing the field strength to lower the gate time unavoidably lowers the overall gate fidelity as well. Nevertheless, the gate fidelity is still above $90\%$ in most of the parameter space. }
}
\label{fig:2D_fidelity_time_Rx}
\end{figure} 

In general, the $CR_X(\phi)$ fidelity is directly proportional to the external magnetic field strength and, therefore, it is inversely proportional to the rotation angle $\phi_k$ in both CPMG and UDD$n$ sequences, which, seeing that in general $\phi_k\gg\tilde{\phi}_k$ (see Eqs.~(\ref{eq:conditional_rotation}-\ref{eq:unconditional_rotation_UDD4})), implies that a higher-fidelity $CR_X(\phi)$ would result in an undesirably long $R_x(\tilde{\phi})$ gate time. Having excessively slow single-qubit gates would hamper any further development that involves nuclear spins in defects as quantum registers or processors. However, the use of weaker external magnetic fields not only improves the nuclear spin selectivity but it also lowers the gate time of the single-qubit $x$-rotation caused by a dynamical decoupling sequence with an off-resonance unit time $t$. Fig.~\ref{fig:2D_fidelity_time_Rx} shows the fidelity and gate time for the CPMG-based single-qubit rotation $R_x(\tfrac{\pi}{2})$, which are calculated for a range of hyperfine parameter values and different magnetic field strengths. We choose to use only the CPMG sequence for the calculations presented in Fig.~\ref{fig:2D_fidelity_time_Rx} because the other sequences (UDD and CPMG+UDD) give similar fidelities but worse gate times. As shown in Fig.~\ref{fig:2D_fidelity_time_Rx}(c,d), a lower magnetic field strength reduces the overall single-qubit $x$-rotation gate time; notwithstanding, a weak magnetic field also reduces the gate fidelity (Fig.~\ref{fig:2D_fidelity_time_Rx}(a,b)). Nevertheless, the gate fidelity is still above 90\% in most of the parameter space under weak magnetic fields.

We next test the spin selectivity of the hybrid protocol. In Fig. \ref{fig:selectivity_hybrid}, we compare the spin selectivity of CPMG and CPMG+UDD4 hybrid  protocols for two nuclear spins with parameters taken from Ref.~\cite{TaminiauPRL2012}. We set the nuclear Larmor frequency equal to $\omega_L/2\pi$=314 kHz,  which is the  setup in Fig. \ref{fig:selectivity_2spins}. We choose to use  $k=3$, the third resonance, for all CPMG and UDD4 pulse sequences, wherein two spin resonance times  are more separated to achieve better spin selectivity. In the totally entangling process of two nuclear spin respectively, we find that the hybrid protocol and the CPMG protocol have the same gate times ($T=t (N_{\text{CPMG}}+N_{\text{UDD4}})$), which are only determined by their sequence iteration numbers $N$: (a) $N_{\text{CPMG}}=9$, (c) $N_{\text{CPMG}}=8,N_{\text{UDD4}}=1$; (b) $N_{\text{CPMG}}=18$, (d) $N_{\text{CPMG}}=17, N_{\text{UDD4}}=1$. The two protocols show similar spin selectivity, which is not surprising, given that the hybrid protocol generally consists of a long sequence \textcolor{black}{(i.e, large $N_{\text{CPMG}}$)} of CPMG pulses, followed by a short sequence \textcolor{black}{(i.e, small $N_{\text{UDD}}$)} of UDD4 pulses.

Now we turn our focus toward the coherence-preserving power of the hybrid sequence CPMG+UDD. The filter function for the hybrid sequence (CPMG)$^{N_{\mathrm{CPMG}}}$+(UDD$n$)$^{N_{\mathrm{UDD}}}$ is a combination of Eqs.~(\ref{eq:filter_function_CPMG^N},~\ref{eq:filter_function_UDDn^N}) and is given by
\begin{equation}\label{eq:filter_function_CPMG+UDD}
\begin{aligned}
    F^{\text{CPMG+UDD\textit{n}}}_{N_{\text{CPMG}},N_{\text{UDD}}}(\omega T)&=\vert  1+(-1)^{2N_{\text{CPMG}}+n N_{\text{UDD}}+1}e^{i\omega T}\\
    &+2\sum_{j_1=1}^{2N_{\text{CPMG}}}(-1)^{j_1}e^{i\delta_{j_1}^{\text{CPMG}}\omega T}\\
    &+2\sum_{l=0}^{N_{\text{UDD}}-1}\sum_{j_2=1}^{n}(-1)^{l n +j_2}e^{i \delta_{l n+j_2}^{\text{UDD}}\omega T}\vert^2,
\end{aligned}
\end{equation}
where $\delta_j^{\text{CPMG}}=(j-1/2)/[(2(N_{\mathrm{CPMG}}+N_{\mathrm{UDD}}))$ and $\delta_{l n+j}^{\mathrm{UDD}}=(l+N_{\mathrm{CPMG}})/(N_{\mathrm{CPMG}}+N_{\mathrm{UDD}})+\sin^2[\pi j/(2n+2)]/(N_{\mathrm{CPMG}}+N_{\mathrm{UDD}})$. In Eq.~\eqref{eq:filter_function_CPMG+UDD} we assume that the order of the full sequence, from left to right, is CPMG first followed by UDD. For alternative orders the fractional pulse locations must be slightly modified.

\begin{figure} [tb]
\includegraphics[width=\linewidth]{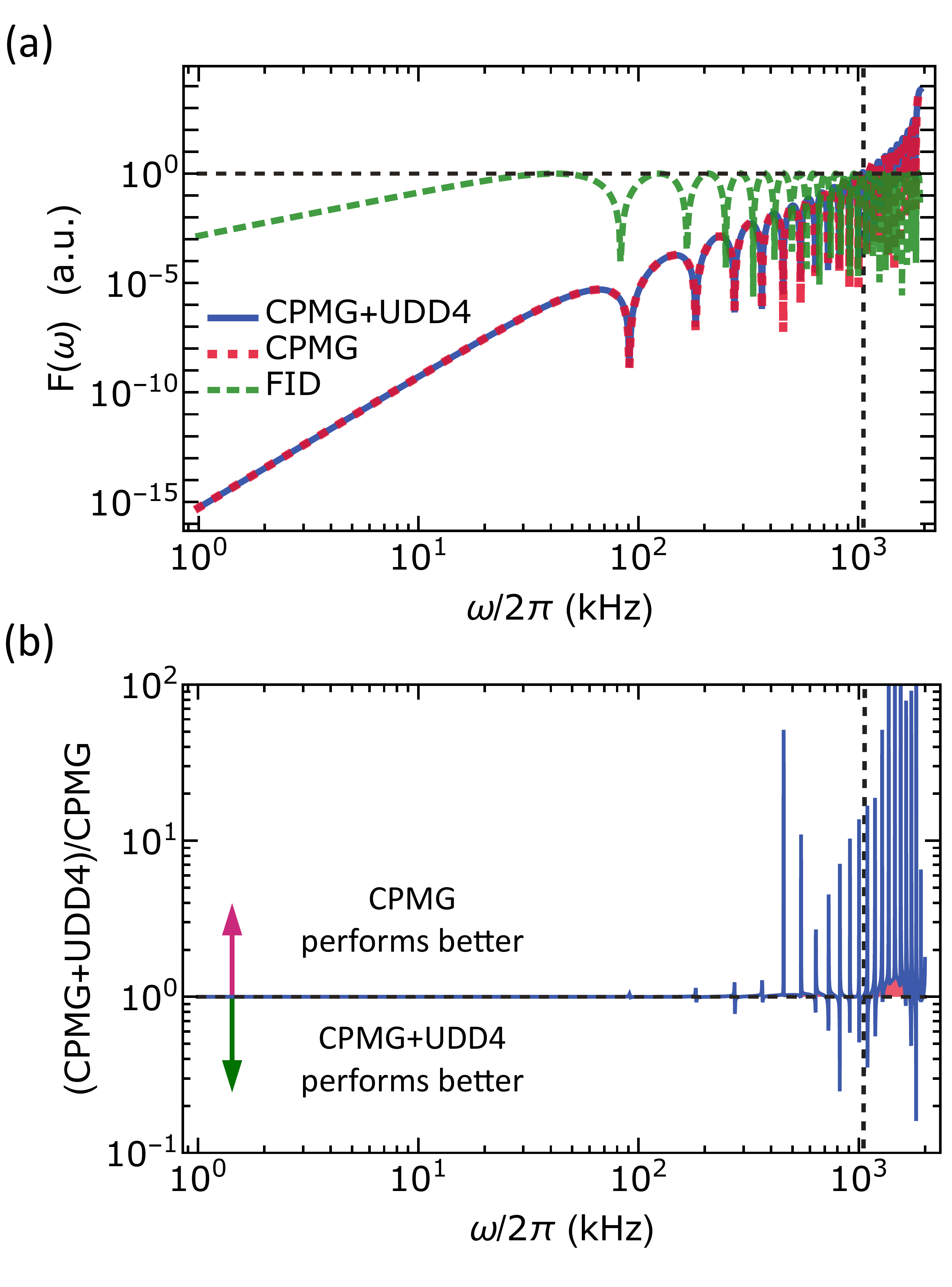} 
\caption{Comparison of the CPMG+UDD4 and CPMG filter functions as functions of noise frequency $\omega/2\pi$. The parameters for the pulse sequences used in the numerical calculations are $\{N_{CPMG}=21,N_{UDD}=2,T=12.01\ \mu\text{s} \}$ for CPMG+UDD and $\{N=21, T=10.97\ \mu\text{s}\}$ for CPMG. Both pulse sequences induce a high-fidelity $CR_x(\tfrac{\pi}{2})$ gate between the electron and target nuclear spin when the latter has the following hyperfine interaction parameters $\{A_{\perp}/2\pi=70\ \text{kHz},A_{\parallel}/2\pi=170\ \text{kHz}\}$. The pulse sequence and hyperfine parameters were extracted from Figs.~\ref{fig:2D_fidelity_time_CNOT}(c,d). For the free-induction decay (FID) process, the electron spin state freely evolves for 12.01 $\mu$s.(a) Numerically calculated filter functions for CPMG+UDD4, CPMG, and free-induction decay (FID). The horizontal dashed black line indicates $F(\omega)=1$. (b) The quotient between the numerical values of CPMG+UDD4 and CPMG. A quotient equal to 1 indicates equal filter functions, a quotient less than 1 corresponds to CPMG+UDD4 outperforming CPMG and vice versa for a quotient greater than 1 (magenta shading). Vertical dashed black lines in both plots mark the frequency value $\omega_1/2\pi\approx 1$ MHz below which noise is effectively suppressed by both pulse sequences.
}
\label{fig:FilterFunction}
\end{figure}

We compare the CPMG+UDD4 and CPMG filter functions in Fig.~\ref{fig:FilterFunction}. These pulse sequences, with parameters $\{N_{CPMG}=21,N_{UDD}=2,T=12.01\ \mu\text{s} \}$ for CPMG+UDD4 and $\{N=21, T=10.97\ \mu\text{s}\}$ for CPMG, induce a high-fidelity $CR_x(\tfrac{\pi}{2})$ gate between the electron and a target nuclear spin with hyperfine parameters $\{A_{\perp}/2\pi=70\ \text{kHz},A_{\parallel}/2\pi=170\ \text{kHz}\}$. The pulse sequences and hyperfine parameters were extracted from Figs.~\ref{fig:2D_fidelity_time_CNOT}(c,d), in which the sequence parameters are optimized to generate high-fidelity $CR_X(\tfrac{\pi}{2})$ gates in the shortest time possible. Figure.~\ref{fig:FilterFunction}(a) shows, apart from the CPMG+UDD4 and CPMG filter functions, the filter function for a free-induction decay (FID) process given by~\cite{Uhrig2008,Cywinski2008} $F^{\mathrm{FID}}(\omega T)=\sin^2(\omega T/2)$. In this process the electron spin state is allowed to freely precess for certain time $T$ ($T=12.01\ \mu \mathrm{s}$ in Fig.~\ref{fig:FilterFunction}(a)) under the effect of a dephasing Hamiltonian which, in an ensemble average, produces a decay in coherence. 
The filter functions of both pulse sequences are, as expected, much smaller than the filter function of FID for low-frequency noise but they get closer to each other with increasing noise frequency until they become equal to or greater than 1 (horizontal dashed black line). The vertical dashed black lines in both Figs.~\ref{fig:FilterFunction}(a) and (b) mark the minimum frequency $\omega_1/2\pi\approx 1$ MHz at which both filter functions are equal to 1. Therefore, for noise frequencies equal to or greater than $\omega_1$, both pulse sequences do not effectively suppress the noise and can even amplify the decoherence. In the same vein, Fig.~\ref{fig:FilterFunction}(b) shows that for noise frequencies less than $\omega_1$ both filter functions perform equivalently (the quotient between filter functions is equal to 1) except for noise frequencies close to $\omega_1$, where CPMG slightly outperforms CPMG+UDD4 (quotient greater than 1). However, the noise spectral density near $\omega_1/2\pi\approx 1$ MHz is already considerably small~\cite{deLange60}. Therefore, the hybrid CPMG and UDD pulse sequences, in comparison to CPMG alone, do not appreciably lower the ability to extend the electron spin coherence time.

\section{Conclusions}\label{sec:Conclusions}

In this work we have introduced a new way of conditionally controlling nuclear spins via UDD and hybrid dynamical decoupling sequences acting on the central electron spin in NV centers. The Uhrig sequences provide flexibility in terms of enhancing nuclear spin selectivity, without increasing gate times. Surprisingly, in this case UDD performs better than CPMG in terms of electron spin coherence protection too. 
\textcolor{black}{The hybrid approach combines CPMG and Uhrig dynamical decoupling sequences, and has the advantage of producing fast entangling two-qubit gates between the electron and target nuclear spins with higher fidelity than what would be obtained with using either Uhrig or CPMG sequences alone. This is because UDD gives small rotation angles, which supplement the rotation induced by CPMG. Thus, hybrid sequences overcome the coarse ‘granularity’ issue of CPMG-induced conditional gates. Even though the hybrid protocol does not improve the spin selectivity over CPMG as the UDD case does, it is an improved tool to implement fast and high fidelity gates when the spin selectivity given by CPMG is satisfactory.}
Moreover, the hybrid sequence retains most of the noise-suppression characteristics of CPMG, as shown by its filter function. In addition, in contrast to other sequences, the hybrid protocol is less restrictive regarding the strength of the external magnetic field, allowing the use of weaker magnetic fields without significantly increasing the overall gate time and, at the same time, giving better spin selectivity. It also allows the use of very strong magnetic fields while reducing the sequence sensitivity to pulse timing error. 

Our results are applicable not only to NV centers but also to similar defect platforms such as the \textcolor{black}{SiV$^{0}$} in diamond~\cite{RosePRB2018} and divacancy centers in SiC \cite{alex2020entanglement,AbramPRL2015,KoehlNature2011,ChristleNatMat2014,FalkNatCom2013}, \textcolor{black}{both of which have a ground state with $S=1$}. The latter is particularly interesting since it has two types of nuclear spins, ${}^{13}_{}\text{C}^{}_{}$ and ${}^{29}_{}\text{Si}^{}_{}$, which can be treated as two independent nuclear spin baths due their negligible interference~\cite{PhysRevB.90.241203}. Overall, our work presents a more versatile way to control weakly coupled nuclear spins via a central electron spin, and thus it is immediately relevant to experiments with existing capabilities in NV centers and similar systems. 

\begin{acknowledgments}
We thank T. Taminiau and Mo Chen  for helpful discussions. This work was supported by the NSF (Grant Numbers 183897 and 1741656).
\end{acknowledgments}

\appendix
\section{Effective CNOT gate}\label{app: A - Effective cnot gate}
Here we explain that the conditional nuclear spin rotation gate $CR_X(\frac{\pi}{2})$ is effectively equivalent to the CNOT gate (CNOT= $\ket{0}\bra{0}\otimes I+\ket{1}\bra{1} \otimes X$) up to single-qubit  gates and a trivial phase. To that end, we start from a nuclear spin coupled to the central electron spin and apply $N$ iterations of decoupling units satisfying the resonance condition. The $N$ is chosen such that the total evolution operator is:
\begin{equation}
\begin{aligned}
  CR_X&\big(\frac{\pi}{2}\big)= \ket{0}\bra{0}\otimes R_X\big(\frac{\pi}{2}\big)+\ket{1}\bra{1}\otimes R_{-X}(\frac{\pi}{2})\\
\end{aligned}
 \end{equation} 
where  $R_X\big(\frac{\pi}{2}\big)=e^{-i\frac{\pi}{4}\sigma_X}$ and $R_{-X}(\frac{\pi}{2})=e^{i\frac{\pi}{4}\sigma_X}$. Since $R^{-1}_X(\frac{\pi}{2}) R_{-X}(\frac{\pi}{2}) = i X$, we can make the following simplification:
\begin{equation}
\begin{aligned}
  CR_X&\big(\frac{\pi}{2}\big)= R_X\big(\frac{\pi}{2}\big)\big(\ket{0}\bra{0}\otimes I+\ket{1}\bra{1}\otimes i X\big) \\
= & \frac{1+i}{\sqrt{2}} R_X\big(\frac{\pi}{2}\big)\bigg(\frac{1-i}{\sqrt{2}}\ket{0}\bra{0}\otimes I+\frac{1+i}{\sqrt{2}}\ket{1}\bra{1}\otimes  X\bigg) \\
= & \frac{1+i}{\sqrt{2}} R^{(E)}_z\big(\frac{\pi}{2}\big)R_X\big(\frac{\pi}{2}\big) \bigg(\ket{0}\bra{0}\otimes I+\ket{1}\bra{1}\otimes  X\bigg).
\end{aligned}
 \end{equation} 
It is evident from the expression that $CR_X(\frac{\pi}{2})$ is equivalent to the  CNOT gate, up to an electron spin state-independent nuclear gate $R_X(\frac{\pi}{2})$, a  $R^{(E)}_z\big(\frac{\pi}{2}\big)$ gate on the electron spin, and an ignorable trivial phase. 

Since $CR_X(\frac{\pi}{2})=\bigg(R^{(E)}_z\big(\frac{\pi}{2}\big) R_X\big(\frac{\pi}{2}\big) \bigg) \text{CNOT}$,  to implement the standard CNOT gate  one should apply the corresponding unconditional nuclear gates and electron spin gate to counteract the $R^{(E)}_z\big(\frac{\pi}{2}\big) R_X\big(\frac{\pi}{2}\big)$.

\section{System Hamiltonian}\label{app: B - System Hamiltonian}
We consider a nuclear spin $^{13}$C that interacts with the central electron spin in the presence of a magnetic field along the $z$ direction. 
The total Hamiltonian, Eq.~\eqref{eq:total_Hamiltonian} in the main text, can be decomposed as the summation of the  following terms: 
\begin{equation}
\begin{aligned}
H_{E}&=-\gamma_e B  S_z+\Delta S^2_z \\
H_{int}& =  \vec{S} \cdot \mathbb{A} \cdot \vec{I}   \\
H_{bath}&=- \gamma_C B I_z,
\end{aligned}
\end{equation}
where $\gamma$ is the gyromagnetic ratio,  $\Delta$ is the zero field splitting, vectorized spin operators $\vec{S},\vec{I}$  contain $x,y,z$ components each, and   $\mathbb{A}$ =$\mathbb{A}_{i,j}$  is the electron-nuclear hyperfine tensor, which contains 9  components for $i,j \in \{x,y,z\}$. 
Since the nuclear spin is not assumed to be a nearest neighbour of the electron spin, $H_{int}$ can be described by the dipole-dipole interaction:
\begin{equation}
\begin{aligned}
H_{int}& = \vec{S} \cdot \mathbb{A} \cdot \vec{I} \\
& =  \vec{S} \cdot \frac{\mu_0\gamma_e\gamma_{ \text{C} }}{4\pi r^3 } (1-  \frac{3\vec{r} \vec{r} }{r^2 } )  \cdot \vec{I} \\
&=	
		\begin{pmatrix}
		S_x \\
		S_y  \\
		S_z  \\
		\end{pmatrix}^T
			\cdot 
		\begin{pmatrix}
		\mathbb{A}_{xx} & \mathbb{A}_{xy} & \mathbb{A}_{xz}\\
		\mathbb{A}_{yx}  &\mathbb{A}_{yy} &\mathbb{A}_{yz} \\
		\mathbb{A}_{zx}  & \mathbb{A}_{zy} &\mathbb{A}_{zz}\\
		\end{pmatrix} 
				\cdot 
		\begin{pmatrix}
		I_x \\
		I_y  \\
		I_z  \\
		\end{pmatrix},		
\end{aligned}	
\end{equation}
where the $\vec{r}$ denotes the displacement vector from the electron to the nucleus.
When we eliminate the transverse components  of  the electron spin, as explained in the main text, we have:
\begin{equation}
	H_{int} = S_z(\mathbb{A}_{zx} I_x+\mathbb{A}_{zy} I_y +\mathbb{A}_{zz}I_z). 
\end{equation}
A proper rotation of the $x$-$y$ plane can reduce the directions down to $\perp$ and $\parallel$  components w.r.t the $z$-axis. 
In this way, the second term in  Eq.~\eqref{eq:single_Hamiltonian} can be  obtained.

\section{Derivation of the analytical expressions for the rotation angles and resonance times for UDD\textit{n}}\label{app:analytical expressions for UDD}
We consider the system formed by the electron spin interacting with a single nuclear spin, whose Hamiltonian is given by Eq.~\eqref{eq:general_H_E-n}. In the main text, we used the coherence function $L(t)$ of the whole system to find the probability of preserving the initial state of the electron, ($\ket{x}=(\ket{0}+\ket{1})/\sqrt{2}$), after a decoupling sequence, which is $P_x=(1+L(t))/2$. An alternative way to find $P_x$ is using the evolution operator of the nuclear spin alone conditioned on the electron spin input states $\ket{0}$ and $\ket{1}$, $U_0$ and $U_1$ respectively. This is the same approach used in Ref.~\cite{TaminiauPRL2012}. Accordingly, the probability of finding the electron in the initial $\ket{x}$ state after the decoupling sequence is $P_x=(1+M)/2$, with $M=\text{Re} \Tr(U_0 U_1^{\dagger})/2$. The nuclear spin evolution operators after a single UDD$n$ decoupling sequence, with $n$ being an even integer, are:
\begin{equation}
\begin{aligned}
U_0&=\prod_{j=0}^{j=n_{even}}\exp\left[-i h_{(1+(-1)^{j+1})/2}\Delta_j(n) \tau \right],\\
U_1&=\prod_{j=0}^{j=n_{even}}\exp\left[-i h_{(1+(-1)^{j})/2}\Delta_j(n) \tau \right],
\end{aligned}    
\end{equation}
where  $\Delta_j(n)$ is defined as
\begin{equation}
\Delta_j(n)=\frac{\sin\left[\frac{\pi(j+1)}{2n+2}\right]^2-\sin\left[\frac{\pi j}{2n+2}\right]^2}{\sin\left[\frac{\pi}{2n+2}\right]^2},
\end{equation}
with $\Delta_0(n)=\Delta_n(n)=1$. The Hamiltonian $h_{(1+(-1)^{j+1})/2}$($ h_{(1+(-1)^{j})/2}$) is either equal to $h_0=\omega_L I_z $ or $h_1=(\omega_L-A_{\parallel})I_z- A_{\perp} I_x$ (see Eq.~\eqref{eq:00h_0+11h_1 Hamiltonian} in the main text). Here $A_{\parallel}\equiv\omega_h\cos(\theta)$ and $A_{\perp}\equiv\omega_h\sin(\theta)$, where $\omega_h$ is the magnitude of the hyperfine interaction and $\theta$ is the angle between the axes of rotation $\vec{\omega}_L$ and $\vec{\omega}_h$. In the absence of hyperfine coupling, the nuclear spin would precess about the axis $\vec{\omega}_L$ with frequency $\omega_L$ (Larmor frequency). Similarly, in the absence of an external magnetic field, the nuclear spin would precess about the axis $\vec{\omega}_h$ with frequency $\omega_h$. 
\begin{figure} [h]
\includegraphics[width=\linewidth]{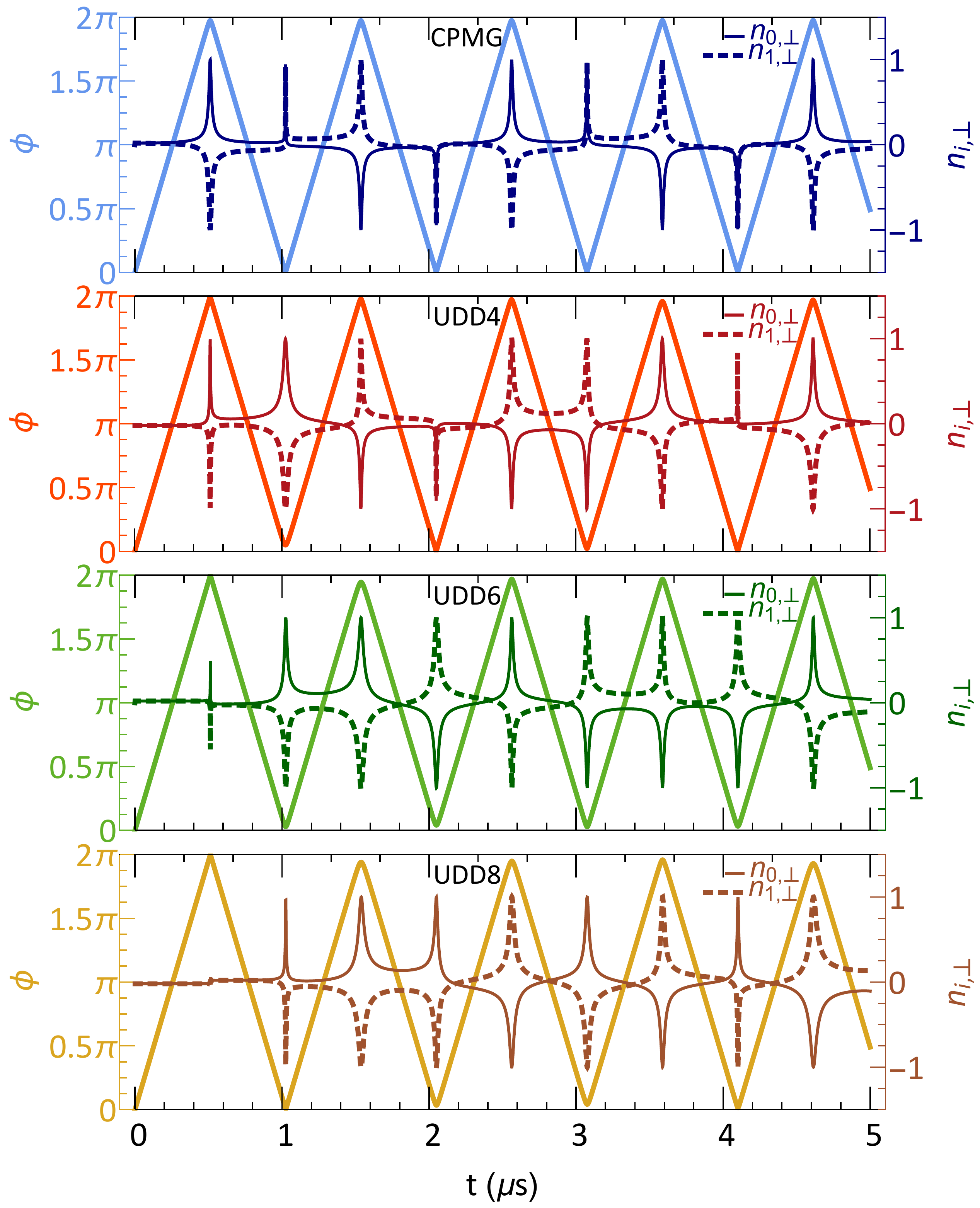} 
\caption{Rotation angle $\phi$ and perpendicular components of the unit axes, $n_{0,\perp}$ and $n_{1,\perp}$, vs the unit sequence time $t$ for different UDD$n$ sequences with even $n$. The values for the system parameters used to make the plots are $\omega_L/2\pi=2$~MHz and $A_{\parallel}/2\pi=A_{\perp}/2\pi=0.1$~MHz. Note that whenever a peak(dip) of $n_{0,\perp}$ coincides with the dip(peak) of $n_{1,\perp}$ the nuclear spin undergoes a conditional rotation.}
\label{fig:UDDevenplots}
\end{figure} 
\begin{figure} [h]
\includegraphics[width=\linewidth]{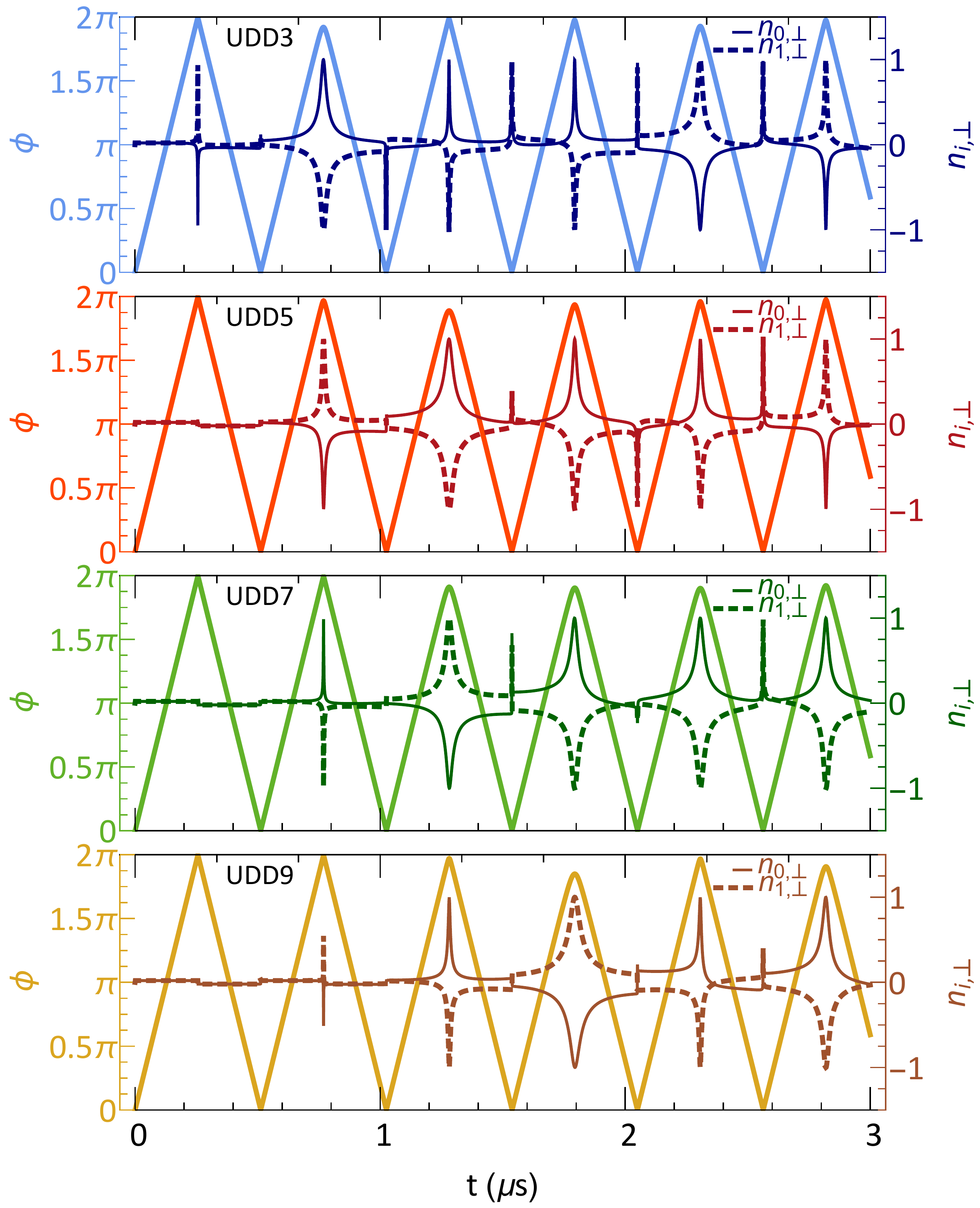} 
\caption{Rotation angle $\phi$ and perpendicular components of the unit axes, $n_{0,\perp}$ and $n_{1,\perp}$, vs the unit sequence time $t$ for different UDD$n$ sequences with odd $n$. We use the same values for the system parameters used in Fig.~\ref{fig:UDDevenplots}. The regions where a peak, nor a dip, is observed but one would otherwise expect to do so, e.g. the region under the first peak of the rotation angle, is due to very sharp processes (and, therefore, quite sensitive to timing imprecision) that were not picked up by the numerical calculations or simply due to the absence of conditional or unconditional $x$-rotations.}
\label{fig:UDDoddplots}
\end{figure} 
\begin{figure*} [!hbt]
\includegraphics[width=\linewidth]{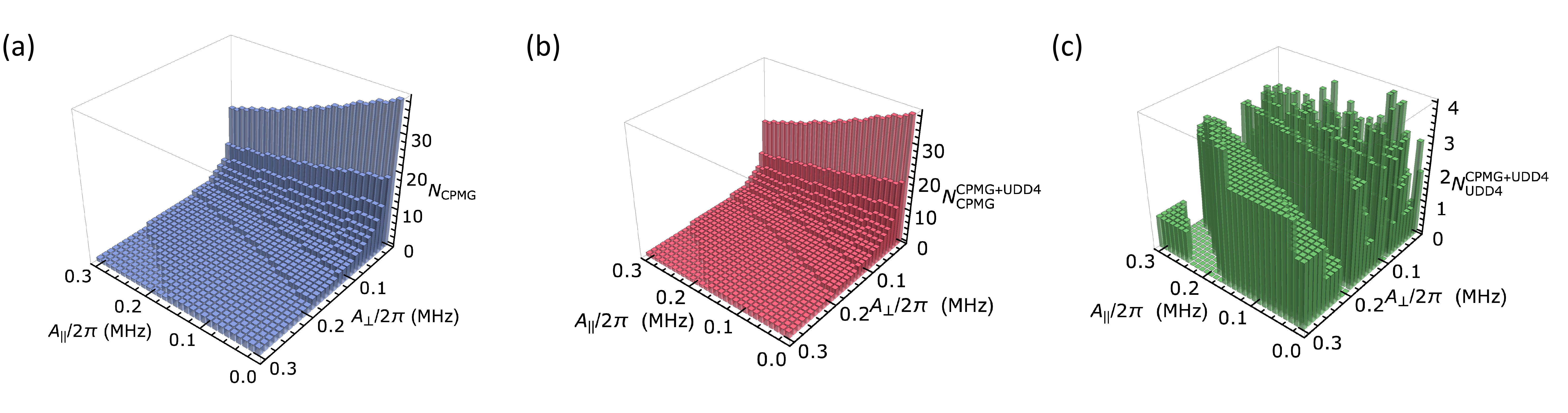} 
\caption{(a) Integer values for the iteration number $N_{\text{CPMG}}$ used in the fidelity plot of Figs.~\ref{fig:2D_fidelity_time_CNOT}(c).(b,c) Integer values for the iteration numbers $N_{\text{CPMG}}^{\text{CPMG+UDD4}}$ and $N_{\text{UDD4}}^{\text{CPMG+UDD4}}$ used in the fidelity plots of Fig.~\ref{fig:2D_fidelity_time_CNOT}(d). The Larmor frequency is set equal to $\omega_L/2\pi=0.5$ MHz.}
\label{fig:N_CPMG_UDD4_wl_2}
\end{figure*} 
\begin{figure*} [!hbt]
\includegraphics[width=\linewidth]{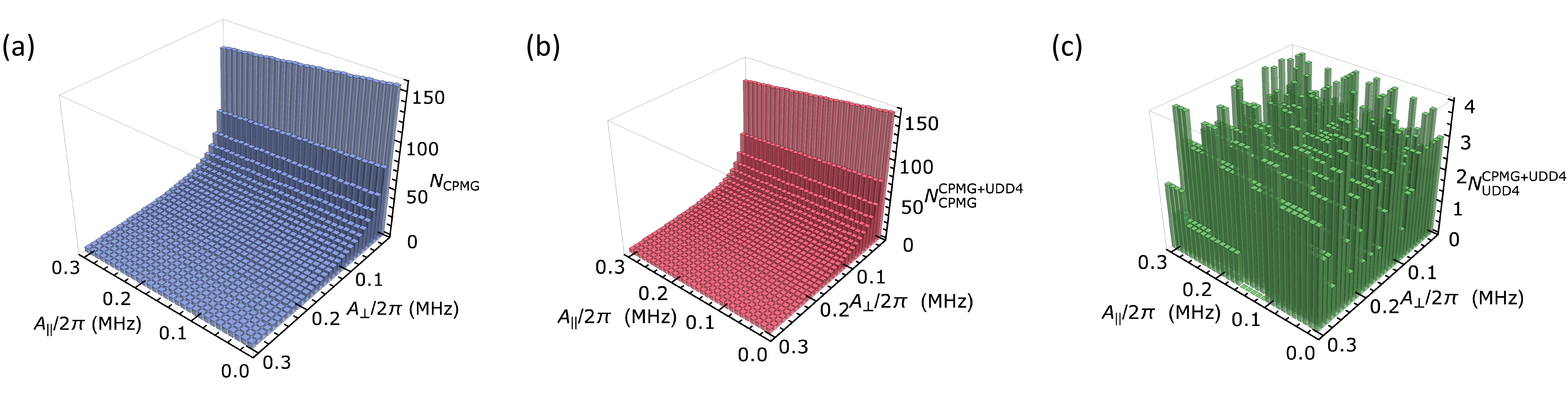} 
\caption{(a) Integer values for the iteration number $N_{\text{CPMG}}$ used in the fidelity plot of Figs.~\ref{fig:2D_fidelity_time_CNOT}(a).(b,c) Integer values for the iteration numbers $N_{\text{CPMG}}^{\text{CPMG+UDD4}}$ and $N_{\text{UDD4}}^{\text{CPMG+UDD4}}$ used in the fidelity plots of Fig.~\ref{fig:2D_fidelity_time_CNOT}(b). The Larmor frequency is set equal to $\omega_L/2\pi=2$ MHz.}
\label{fig:N_CPMG_UDD4_wl_8}
\end{figure*} 

Note that for odd $n$ the operators $U_0$ and $U_1$ do not start and end with the same single evolution operator as is the case for even $n$, a required symmetry that implies that the electron spin returns to its initial state after the decoupling sequence. Therefore, as stated in the main text, for UUD$n$ with  odd $n$ the basic unit sequence must be a combination of two single UDD$n$ sequences, giving the following nuclear spin evolution operators
\begin{equation}
\begin{aligned}
U_0&=\prod_{j=0}^{j=n_{odd}}\exp\left[-i h_{(1+(-1)^{j+1})/2}\Delta_j(n) \tau \right]\\
&\times \prod_{j=0}^{j=n_{odd}}\exp\left[-i h_{(1+(-1)^{j})/2}\Delta_j(n) \tau \right] ,\\
U_1&=\prod_{j=0}^{j=n_{odd}}\exp\left[-i h_{(1+(-1)^{j})/2}\Delta_j(n) \tau \right]\\
&\times\prod_{j=0}^{j=n_{odd}}\exp\left[-i h_{(1+(-1)^{j+1})/2}\Delta_j(n) \tau \right].
\end{aligned}    
\end{equation}

Given that the operators $U_0$ and $U_1$ belong to the SU(2) group, they can be expressed as rotations by an angle $\phi$ around a unit axis $\vec{n}_i$, this is
\begin{equation}\label{eq:single-qubit_rotations_U0_U1}
\begin{aligned}
U_0&=\exp\left[-i \frac{\phi}{2} \vec{\sigma}\cdot\vec{n}_0 \right],\\
U_1&=\exp\left[-i \frac{\phi}{2} \vec{\sigma}\cdot\vec{n}_1 \right],
\end{aligned}
\end{equation}
where $\vec{\sigma}$ is the Pauli vector. Note that the angle of rotation $\phi$ is independent of the electron spin input state because $\Tr(U_0)/2=\Tr(U_1)/2=\cos[\phi/2]$. Using the expressions in Eq.~\eqref{eq:single-qubit_rotations_U0_U1} we obtain~\cite{TaminiauPRL2012} 
\begin{equation}\label{eq:M_equation}
    M=1-(1-\vec{n}_0\cdot\vec{n}_1)\sin[\phi/2]^2,
\end{equation}
which implies that the probability $P_x$ that the initial electron spin state $\ket{x}$ is preserved after the decoupling sequence is equal to 1 if the unit axes $\vec{n}_0$ and $\vec{n}_1$ are parallel, i.e. $\vec{n}_0\cdot\vec{n}_1=1$. On the other hand, for antiparallel axes ($\vec{n}_0\cdot\vec{n}_1=-1$) $P_x$ is effectively the furthest from 1 (the exact value would depend on the magnitude of $\phi$), and thus the electron spin is coupled to the nuclear spin. Note that the electron and nuclear spins are maximally entangled when $\vec{n}_0\cdot\vec{n}_1=-1$ and $\phi=\pi/2$ (or $\phi=3\pi/2$), and thus $P_x=0.5$. 

Approximate analytical expressions for the resonance time $t$ and angle of rotation $\phi$ can be found for any UDD$n$ sequence (including CPMG, i.e. UDD2) using Eq.~\eqref{eq:M_equation} and $\cos[\phi/2]=\Tr(U_0)/2=\Tr(U_1)/2$. First, assuming a high external magnetic field such that $\omega_L\gg\omega_h$, we perform a Taylor series expansion in terms of $\omega_h/\omega_L$ on both Eq.~\eqref{eq:M_equation} and $\cos[\phi/2]=\Tr(U_0)/2$. We only need to keep terms up to first order to find the approximate analytical expression for the resonance time $t$. We do so by first finding an approximate expression up to first order for the angle $\phi$ using equation $\cos[\phi/2]=\Tr(U_0)/2$. And  then plugging it in Eq.~\eqref{eq:M_equation}, where $M$ reduces to 1 in this first-order approximation and $\vec{n}_0\cdot\vec{n}_1$ is set to -1 to get the interval time $\tau$ needed to implement conditional rotations on the nuclear spin. After obtaining an expression for $\tau$, the unit sequence time $t$ for any UDD$n$ sequence with even or odd $n$ is given by
\begin{equation}
    t=\tau\left(2+\sum_{j=1}^{j=n-1}\tfrac{\sin\left[\frac{\pi(j+1)}{2n+2}\right]^2-\sin\left[\frac{\pi j}{2n+2}\right]^2}{\sin\left[\frac{\pi}{2n+2}\right]^2}\right).
\end{equation}
Finally the approximate analytical expression for the rotation angle $\phi$ can be obtained by plugging the previously found interval time $\tau$ into the Taylor series expansion of $\cos[\phi/2]=\Tr(U_0)/2$ but now we keep terms up to second order. We follow a similar procedure for time $\tilde{t}$ and rotation angle $\tilde{\phi}$ corresponding to unconditional rotations. Analytical expressions for the resonance times and rotation angles for UDD$n$ sequences with $n\geq 5$ are not simple enough to report them here. Moreover, as shown in Figs.~\ref{fig:UDDevenplots} and~\ref{fig:UDDoddplots}, it is not trivial to find a pattern that identifies the $t$ values that produce unconditional rotations in UDD5 and beyond. 

\section{Number of iterations of the sequences used in Fig.~\ref{fig:2D_fidelity_time_CNOT}}\label{appendix:N_values}

Figures~\ref{fig:N_CPMG_UDD4_wl_2} and~\ref{fig:N_CPMG_UDD4_wl_8} show the integer values for the iteration numbers $N_i$ used in the fidelity plots shown in Fig.~\ref{fig:2D_fidelity_time_CNOT} for Larmor frequencies $\omega_L/2\pi=0.5$ MHz and $\omega_L/2\pi=2$ MHz, respectively.

\section{Comparison of UDD4 and CPMG over three nuclear spins }\label{appendix:selectivity_3_spin}
\begin{figure*} [!hbt]
\includegraphics[width=\linewidth]{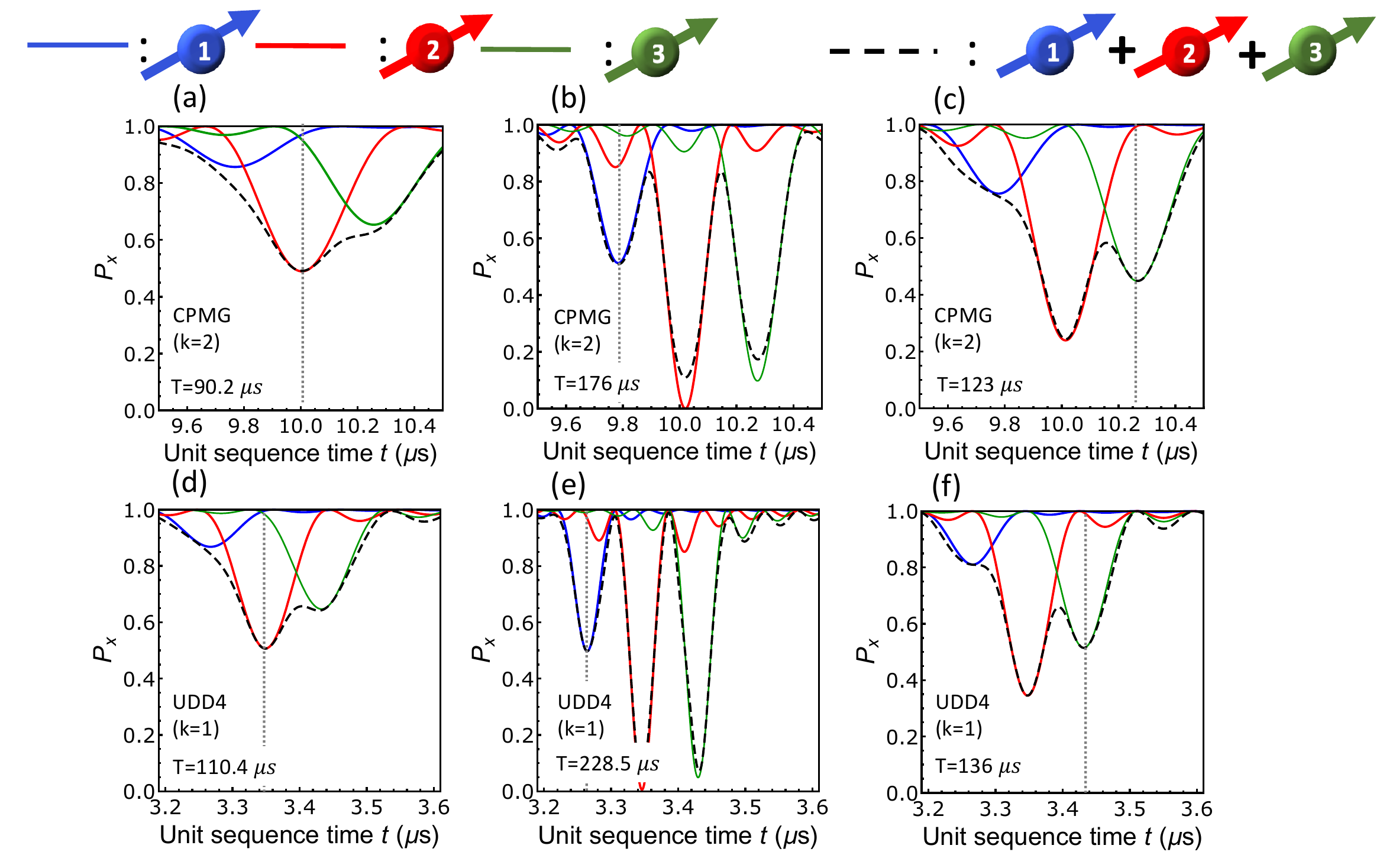} 
\caption{Comparison of UDD4 and CPMG for the case of three nuclear spins. Like in Fig.~\ref{fig:selectivity_2spins}, here we plot the probability $P_x$ of preserving the electron’s initial $\ket{x}$ state after the (a,b,c) CPMG protocol and the (d,e,f) UDD4 protocol.}
\label{fig:selectivity_3_spins_append}
\end{figure*}

Here we compare the spin selectivity power between UDD4 and CPMG protocols, for the case of three nuclear spins. From Fig.~\ref{fig:selectivity_3_spins_append}, we can see that the independent nuclei approximation continues to be valid for three nuclear spins. We note that Delft group has successfully demonstrated the implementation of quantum error correction \cite{Taminiau:2014aa}, contextuality test \cite{vanDamPRL2019}, and decoherence free subspace encoding \cite{ReisererPRX2016} based on the  selective control of several nuclear spins using  dynamical decoupling. Based on the recent results on many nuclear spin manipulation ($>$5) by the Delft group \cite{BradleyPRX2019,Abobeih2018,AbobeihNature2019}, we believe that the independent nuclei approximation will continue to hold as the number of nuclear spin increases. 

\clearpage 

\bibliography{references}

\end{document}